\documentclass[aps,prx,twocolumn,notitlepage,showpacs,floatfix]{revtex4-2} 
\usepackage[utf8]{inputenc}
\usepackage{morefloats}
\usepackage[dvips]{graphicx,color}
\usepackage{transparent}
\usepackage[dvipsnames]{xcolor}
\usepackage{epsfig,amsbsy}
\usepackage{amsmath,amsfonts,amsthm,amssymb}
\usepackage{url}
\usepackage{verbatim}
\usepackage[colorlinks=true,allcolors=blue]{hyperref}
\usepackage{array}
\usepackage{multirow}
\usepackage[normalem]{ulem}
\usepackage{braket} 
\usepackage{dsfont}
\usepackage{natbib}
\usepackage{multibib}
\usepackage{bbold}
\usepackage{hyperref}
\usepackage{makecell}
\setcellgapes{3pt}
\usepackage{physics}
\usepackage{txfonts}
\usepackage{titlesec}

\renewcommand{\vec}[1]{\ensuremath{\boldsymbol{#1}}}
\newcommand{\aG}{\alpha_\text{G}}
\newcommand{\NoSp}{\mathcal{S}}

\begin{document}

\title{Long-distance coupling of spin qubits via topological magnons}
\author{Bence Het\'enyi}
\affiliation{Department of Physics, University of Basel, Klingelbergstrasse 82, CH-4056 Basel, Switzerland}
\author{Alexander Mook}
\affiliation{Department of Physics, University of Basel, Klingelbergstrasse 82, CH-4056 Basel, Switzerland}
\author{Jelena Klinovaja}
\affiliation{Department of Physics, University of Basel, Klingelbergstrasse 82, CH-4056 Basel, Switzerland}
\author{Daniel Loss}
\affiliation{Department of Physics, University of Basel, Klingelbergstrasse 82, CH-4056 Basel, Switzerland} 
\date{\today}

\begin{abstract}
We consider two distant spin qubits in quantum dots, both coupled to a two-dimensional topological ferromagnet hosting  chiral  magnon edge states at the boundary. The chiral magnon is used to mediate entanglement between the spin qubits, realizing a fundamental building block of scalable quantum computing architectures: a long-distance two-qubit gate. Previous proposals for long-distance coupling with magnons involved off-resonant coupling, where the detuning of the spin-qubit frequency from the magnonic band edge provides protection against spontaneous relaxation. The topological magnon mode, on the other hand, lies in-between two magnonic bands far away from any bulk magnon resonances, facilitating strong and highly tuneable coupling between the two spin qubits. Even though the coupling between the qubit and the chiral magnon is resonant for a wide range of qubit splittings, we find that the magnon-induced qubit relaxation is vastly suppressed if the coupling between the qubit and the ferromagnet is antiferromagnetic. A fast and high-fidelity long-distance coupling protocol is presented capable of achieving spin-qubit entanglement over micrometer distances with $1\,$MHz gate speed and up to $99.9\%$ fidelities. The resulting spin-qubit entanglement may be used as a probe for the long-sought detection of topological edge magnons.
\end{abstract}

\maketitle

\section{Introduction}

Along the journey towards universal quantum computing several of the milestones~\cite{loss1998quantum} have already been reached, such as single-qubit gates with long coherence times and fast readout as well as short-ranged two-qubit gates in multiple platforms~\cite{kok2007linear,qiang2018large,ballance2016high,waldherr2014quantum,dolde2014high,barends2014superconducting,arute2019quantum,veldhorst2015two,watson2018programmable,he2019two,hendrickx2021four,alfieri2022nanomaterials}. Universality, on the other hand, requires coherent logical qubits, that can be achieved in large-scale quantum computers by means of quantum error correction~\cite{fowler2012surface,wootton2012high}. Owing to the highly developed semiconductor industry qubits defined in semiconductor quantum dots (QDs)~\cite{kloeffel2013prospects,chatterjee2021semiconductor,burkard2021semiconductor} are increasingly believed to be an exceptionally potent candidate for the long term goal: scalable quantum computers. The challenge incorporates the improvement of single- and two-qubit gate performance as well as the management of the corresponding control electronics~\cite{veldhorst2017silicon,vandersypen2017interfacing,li2018crossbar}. Leveraging the industry-standard fabrication techniques, Ref.~\cite{vandersypen2017interfacing} proposed to accommodate elements of the control electronics on the same chip by arranging small dense qubit arrays and local control electronics in a checkerboard pattern, where the qubit arrays are connected via long-range qubit couplers. For such architectures having means to create entanglement over large distances ($\gtrsim 1\,\mu$m) would be highly desirable. 

Long-range entanglement of spin qubits is realizable using a variety of mediators~\cite{burkard2021semiconductor} such as floating gates \cite{trifunovic2012long,szumniak2015long}, microwave cavities \cite{kloeffel2013circuit}, superconducting resonators \cite{nigg2017superconducting,brojans2020resonant,harvey2022coherent} or spin shuttling \cite{mcneil2011demand,boter2019sparse,yoneda2021coherent}. While the fidelity of the aforementioned protocols may be limited by charge noise, magnetic insulators are versatile platforms to create entanglement among distant spins with low dissipation and no heat generation due to Joule heating \cite{Barman_2021}, Furthermore, the coupling to magnons does not require spin-orbit interaction (SOI). In such systems the effective coupling between spin qubits can be established using ferromagnetic (FM) magnons  \cite{trifunovic2013long, Candido_2020, Skogvoll2021}, antiferromagnetic domain walls \cite{flebus2019entangling} or magnon waveguides \cite{fukami2021opportunities}.  An other promising approach to mitigate dissipation is to couple spin qubits via topological edge states in quantum Hall systems \cite{yang2016long,elman2017long,wagner2019driven,bosco2019transmission}.

\begin{figure*}[th]
    \includegraphics[width= 0.95\textwidth]{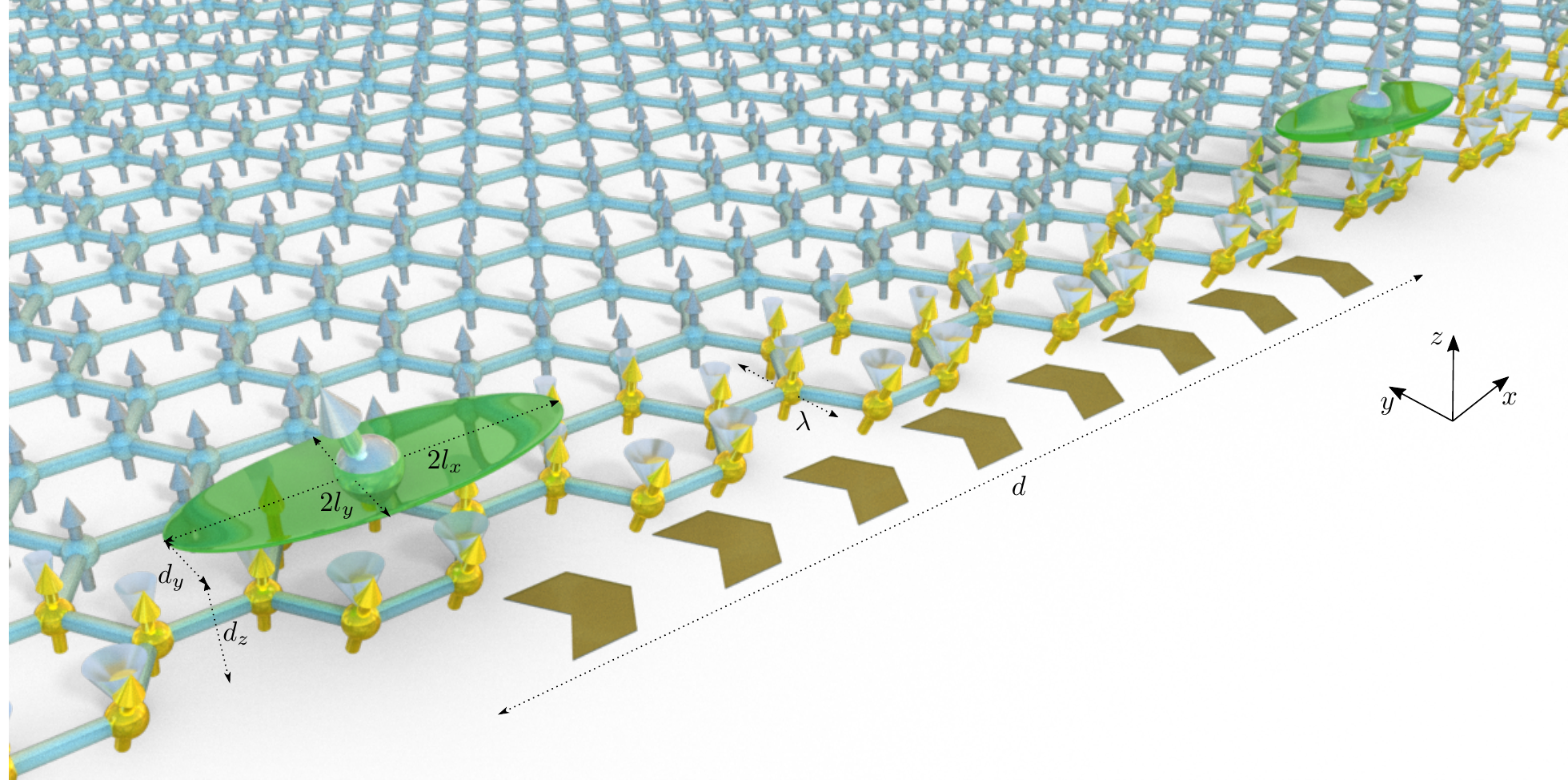}
    \caption{Schematic setup for the long-distance spin-qubit entanglement mediated by chiral magnons in a topological ferromagnet. The light blue honeycomb lattice represents the ferromagnet with the arrows indicating the ground state spin polarization. The armchair edge of the ferromagnet hosts the chiral magnon mode propagating along the positive $x$ direction, indicated by the canted edge spins (gold).  The spin qubits (silver arrow embedded in a green ellipse) are lying in a parallel plane close to the FM lattice (qubit layer is not shown explicitly), located near the edge of the magnet.}
    \label{fig:system}
\end{figure*}

Herein, as shown in Fig.~\ref{fig:system}, we bring together topological excitations, magnets, and spin qubits by studying long-distance entanglement mediated by topological magnons. The latter are examples of bosonic topological spin excitations above topologically trivial magnetic ground states. Topological chiral magnons are predicted to exist in a large variety of magnetic systems, ranging from FM \cite{Meier2003, Katsura2010, Shindou13, Shindou13b, Zhang2013, Mook2014, Shindou14, Mook2014edge, Nakata2017} and antiferromagnets \cite{Nakata2017AFM, Mook2018duality, Mook2019Coplanar} to skyrmion crystals \cite{Hoogdalem2013, RoldanMolina2016, Garst2017, Diaz2019AFM, Kim2019SkyrmionHallFerri, diaz2020chiral}, and from two-dimensional to three-dimensional systems \cite{Li2016Weyl, Mook2016Weyl, mook2021chiral}. Being nonconserved bosons, chiral edge (or interface \cite{Shindou13, Mook2015waveguide, Mook2015interfaces}) magnons exist within topological spectral gaps at finite frequencies, typically between a few GHz in (artificially manufactured or self-organized) topological magnonic crystals \cite{Hoogdalem2013, Shindou13, Shindou13b, Shindou14, Xu2016, RoldanMolina2016, Garst2017, Iacocca2017, Li2018, Diaz2019AFM, Li2018magcrys, Mellado_2022} up to several THz in magnetic compounds. Examples for the latter are Cu(1,3-benzenedicarboxylate) \cite{Chisnell2015}, CrI$_3$ \cite{Chen2018CrI3}, CrSiTe$_3$, and CrGeTe$_3$ \cite{Zhu2021}. For recent reviews on topological magnons, see Refs.~\cite{Malki_2020, Li_2021, McClarty_2021, Wang_2021}.

Once the qubit is brought into proximity to the magnet's edge (or interface) and its frequency is tuned within the topological magnon gap, the qubit is only resonant with the chiral edge mode. Coupling the qubit to the FM leads to an emission of a \emph{physical} unidirectionally propagating magnon well localized to the edge of the sample. This magnon can be reabsorbed by the second qubit thereby mediating entanglement between the qubits. This nonreciprocal coupling protocol can be exceptionally fast ($\sim 1\,$GHz) and we find high gate fidelities when the inter-qubit distance is well below the magnon mean free path, that is to say, well below $1\,\mu$m.

Importantly, we report a coupling regime that drastically outperforms the aforementioned protocol. If the two qubits are coupled simultaneously with the FM (antiferromagnetically), a \emph{virtual} chiral magnon-mediated process arises, which is proportional to the direct exchange coupling, with the decoherence rates being suppressed by the smallness of the dipole-dipole interaction. In this regime, fidelities of $99.9\%$ of $1\,$MHz two-qubit gates can be achieved even at distances comparable with the magnon mean free path.

The remainder of this work is structured as follows: in Sec.~\ref{sec:FM} the model of a two-dimensional topological FM in nanoribbon geometry is presented and its chiral edge magnons characterized. In Secs.~\ref{sec:QD}-\ref{sec:FMQDcoup}, we consider two planar QDs residing in an adjacent non-magnetic layer, coupled by both direct exchange and dipole-dipole interaction to an armchair edge of the FM.  In Secs.~\ref{sec:effQDQDcoup}-\ref{sec:decoh}, we identify a coupling regime with antiferromagnetic exchange coupling between the QD and the FM. In this scenario the qubit relaxation is orders of magnitude slower than the effective coupling. In Secs.~\ref{sec:chiral}-\ref{sec:exanal}, we present the results of the corresponding numerical study, which we show to agree well with our analytical estimates. Finally, we consider the opposite (ferromagnetic) coupling regime in Sec.~\ref{sec:transd} for which the resonant coupling together with the chiral propagation of the magnon facilitates qubit entanglement via the exchange of a physical magnon. After a discussion in Sec.~\ref{sec:discussion}, we conclude in Sec.~\ref{sec:conclusion}. Several Appendices provide more detailed information.

\section{Theory}
\subsection{Model of the topological ferromagnet}
\label{sec:FM}

We consider a two-dimensional honeycomb lattice, as shown in Fig.~\ref{fig:system}, with each lattice site---indexed by $i$---hosting a localized spin operator $\vec{S}_i$. Nearest neighbors interact via ferromagnetic Heisenberg exchange interaction, $J>0$, and next-nearest neighbors are coupled via Dzyaloshinsky-Moriya interaction (DMI) \cite{Dzyaloshinsky58,Moriya60}, originating from spin-orbit interaction. The spin Hamiltonian of the FM thus reads as
\begin{align}
    H_\text{FM} = -\frac{J}{2} \sum_{ \langle i,j \rangle} \vec{S}_i \cdot \vec{S}_j  +\frac D 2 \sum_{ \langle  \langle i,j  \rangle  \rangle } \nu_{ij}\, \vec{\hat z} \cdot \left( \vec{S}_i \times \vec{S}_j \right) + H_\text{ani},
    \label{eq:HFM}
\end{align}
where $\nu_{ij} =-\nu_{ji} = \pm 1$ depending on the relative position of sites $i$ and $j$. Here, we adopt the convention that $\nu_{ij} = +1$, if the bond from site $i$ to site $j$ points in anticlockwise direction as seen from the respective hexagon. We also added an anisotropy term $H_\text{ani}$ in Eq.~\eqref{eq:HFM} that gaps out the Goldstone mode by creating a spin-wave gap. Since its microscopic origin is of no further relevance, we model the anisotropy by a built-in magnetic field, $H_\text{ani} = - \Delta_\text{F} \sum_i S^z_i$, into which potential external fields may be absorbed as well. Then, $\Delta_\text{F}$ comprises the energy of the uniform ferromagnetic resonance.

Spin Hamiltonian \eqref{eq:HFM} is well-studied in the context of topological magnons as it realizes the magnonic version of the Haldane model \cite{Haldane1988}, as shown in Ref.~\cite{Owerre2016a}. Here, we do not repeat the derivation but only summarize the most important aspects crucial for the coupling of spin qubits.
Assuming that the spins in the ground state are pointing in the positive $z$ direction, we perform a Holstein-Primakoff transformation \cite{Holstein1940}. To lowest order in the $1/S$ expansion the spin operators are expressed as
\begin{subequations}
\begin{align}
    S^x_{i} &\approx \sqrt{\frac{S}{2}} \left( a^{}_i+a^\dagger_i \right)\, 
    ,
    \label{eq:sx}
    \\
    S^y_{i} &\approx -\mathrm{i}\sqrt{\frac{S}{2}} \left(a^{}_i-a^\dagger_i \right)\, 
    ,
    \label{eq:sy}
    \\
    S^z_{i} &= S- a^\dagger_i a^{}_i\, ,
    \label{eq:sz}
\end{align}
\end{subequations}
where $a_i^\dagger$ and $a^{}_i$ are bosonic creation and annihilation operators, respectively, and $S$ is the spin quantum number. By plugging Eqs.~\eqref{eq:sx}-\eqref{eq:sz} into Eq.~\eqref{eq:HFM}, the spin Hamiltonian can be expanded in bosonic operators. In the harmonic approximation, only the bilinear piece is retained and found to constitute the bosonic equivalent of the Haldane model. Both nearest-neighbor hopping and onsite potentials are proportional to $JS$. The time-reversal symmetry breaking complex next-nearest neighbor hopping is brought about by DMI and, hence, $\propto D S$\cite{pantaleon2017analytical}. The latter causes a topologically nontrivial opening of a band gap that---according to the bulk-boundary correspondence \cite{hatsugaiChernNumberEdge1993, hatsugaiEdgeStatesInteger1993}---supports a chiral magnonic edge mode. 

\begin{figure}
    \includegraphics[width= \columnwidth]{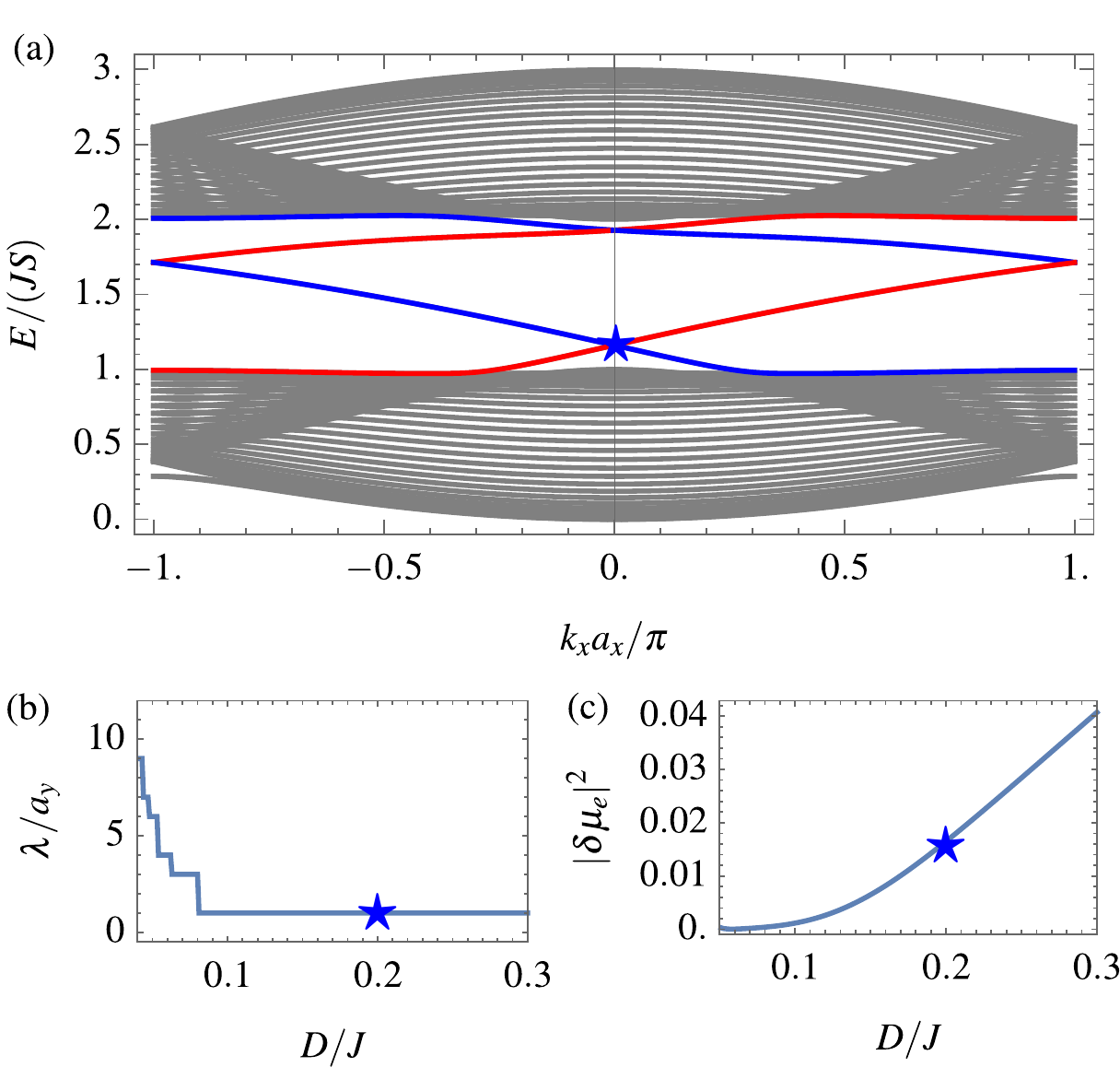}
    \caption{ (a) Magnon spectrum of a honeycomb-lattice ferromagnetic nanoribbon with armchair termination, $D = 0.2\, J$, and $N_y = 20$ unit cells in the $y$ direction. Left (right) localized edge states are shown in blue (red). (b) Localization length $\lambda$ of the left localized edge mode [denoted by a blue star on (a)] as a function of DMI strength. (c) Dynamic magnetic moment of the left localized edge mode as a function of DMI strength.}
    \label{fig:spectrum}
\end{figure}

In the rest of this work, we consider a two-dimensional FM in nanoribbon (or ``slab'') geometry, infinite along the $x$ direction, with armchair termination in the $y$ direction 
\footnote{We have chosen armchair rather than zigzag termination because the latter does not support a chiral edge mode at zero momentum, a property that turns out to be crucial to couple the edge mode to a QD.}. 
The elementary unit cell of size $a_x \times a_y$ contains four atoms (where $a_x = \sqrt 3 a$ and $a_y = a$), and the slab consists of $N_y$ unit cells in the $y$ direction. Using periodic boundary conditions in the $x$ direction, the momentum $k_x \in [ -\tfrac{\pi}{a_x},\tfrac{\pi}{a_x} )$, is a good quantum number and the eigenvalue equation for a given $k_x$ reads as
\begin{align}
    \hat{\vec H}_\text{FM} (k_x) \vec{\varphi}_{k_x,n}   = \varepsilon_{k_x,n} \vec{\varphi}_{k_x,n}\, ,
    \label{eq:HFMkx}
\end{align}
where $n$ is the band index running from $1$ to $4N_y$, where $4N_y$ is the total number of spins in the nanoribbon unit cell. Here, $\hat{\vec H}_\text{FM} (k_x)$ is the linear spin-wave matrix, and $\vec{\varphi}_{k_x,n}$ an eigenvector with eigenvalue $\varepsilon_{k_x,n}$. The eigenvectors satisfy the usual normalization condition, i.e., $\sum_{y_i,\mu} |\varphi^\mu_{k_x,n} (y_i)|^2 = 1$, where $y_i \in [1,N_y]$ is the index of the armchair unit cell and $\mu \in [1,4]$ a basis site within the armchair unit cell. Furthermore, the eigenvectors are related to the spin waves via $S_i^+ \approx \sqrt{2S} a_i = \sqrt{2S/N_x} \sum_{k_x} \mathrm{e}^{-\mathrm{i} k_x x_i} \sum_n\varphi^{\mu_i}_{k_x,n} (y_i) a_{k_x,n}$, where the second equality defines the annihilation operator of the magnonic eigenmode $(k_x,n)$, with $N_x$ being the number of unit cells in the $x$ direction (see App.~\ref{app:conventions} for further conventions). The spectrum $\varepsilon_{k_x,n}$ of such a ferromagnetic slab is shown in Fig.~\ref{fig:spectrum}(a), where the left- and right-propagating chiral edge modes are highlighted in blue and red, respectively. For the numerical results to follow the parameters of the FM slab are listed in Tab.~\ref{tab:FM}, unless otherwise specified.

Importantly, and in contrast to the electronic Haldane model, the chiral mode is not ``particle-hole'' symmetric, i.e., its energy is not symmetric with respect to the gap. This is due to missing nearest neighbors at the edges, resulting in a reduction of energy for the edge modes \cite{Pershoguba2018, Pantaleon2018edge, PANTALEON2018PB}. This edge effect does not affect topological protection because the existence of a chiral mode is still dictated by the nontrivial topology of the bulk. However, it does affect other properties of the chiral edge mode that are related to the edge mode's eigenvector $\vec{\varphi}_{k_x, \text{e}}$ (subscript ``e'' for edge mode) and, as we show, crucial for the spin-qubit coupling. These properties are (i) the edge mode localization length $\lambda = b a_y$, with $b$ being the largest integer for which $ \sum_{y_i = 1}^{b} \sum_{\mu=1}^4 |\varphi^{\mu}_{0, \text{e}} (y_i)|^2\leq 1\!-\!1/e$, and (ii) the edge mode dynamic magnetic moment, which reads as 
\begin{align}
    \delta \mu_\text{e} =  \sum_{y_i = 1}^{N_y} \sum_{\mu=1}^4 \varphi^{\mu}_{0, \text{e}} (y_i) 
\end{align}
for the left edge. These quantities are shown as a function of $D/J$ in Figs.~\ref{fig:spectrum}(b) and (c), respectively. In the following, we work at $D/J = 0.2$, which ensures that the edge state is well localized within the unit cell width, that is to say, $\lambda \sim a_y$.

Later on, we need the transversal spin susceptibility of the topological ferromagnet. In the time domain, we may write it as
\begin{align}
    \chi^\perp_{nm} (t,k_x) \equiv -\mathrm{i}\theta (t) \langle [S_{-k_x,n}^- (t),S^+_{k_x,m} (0)] \rangle \, ,
\end{align}
where $S^-_{-k_x,n} (t) \equiv \sqrt{2S} \mathrm{e}^{\mathrm{i} \varepsilon_{k_x,n} t/\hbar} a^\dagger_{k_x,n}$ encompasses the dynamics associated with the $n$th magnon normal mode in the linear spin-wave approximation. In frequency space, we may rewrite it as
\begin{align}
    \chi^\perp_{nm} (\omega,k_x) = - 2S \frac{\delta_{nm}}{\varepsilon_{k_x,n} (1+\mathrm{i} \aG)-\hbar \omega} \, ,
\label{eq:chinmkx}
\end{align} 
where $\aG$ is the dimensionless Gilbert damping coefficient \cite{Karenowska2016}. This phenomenological constant accounts for the ubiquitous magnetization damping processes without specifying microscopic origins. It brings about a finite spectral broadening $\propto \aG \varepsilon_{k_x,n}$ of the magnon line width proportional to the magnon energy \cite{spinwaves2009}. In high-quality magnetic insulators at low temperatures, as considered here, $\aG \ll 1$, because metallic Stoner excitations and Landau damping are absent, defect scattering is minimized, magnon-magnon scattering is frozen out, and magnon-phonon scattering suppressed. We take $\aG = 10^{-4}$ throughout, a value found, for example, in sub-micrometer yttrium iron garnet films \cite{Dubs_2017}. 

\begin{table}
\caption{Characteristic parameters of the topological ferromagnetic slab assumed for the numerical calculations throughout this work. The slab is assumed to be periodic in the $x$ direction and has armchair termination in the $y$ direction.}
\label{tab:FM}
\begin{tabular}{l c c}
\toprule
Parameter & Symbol & Numerical value \\ 
\cline{1-3}
Exchange coupling & $J$ & $1\,$meV \\ 
DMI & $D$ & $0.2\,$meV \\
Spin quantum nuber & $S$ & 3/2 \\
$g$-factor & $g$ & $2$ \\ 
Ferromagnetic resonance & $\Delta_F$ & $50\,\mu$eV\\
Gilbert damping & $\aG$ & $10^{-4}$\\
Next-nearest-neighbor distance  & $a$ & $1\,$nm \\ 
Slab width & $L_y$ & $20\, $nm \\ 
\botrule
\end{tabular} 
\end{table}

\subsection{Model and requirements for the spin qubits}
\label{sec:QD}
We assume that the spin qubits are defined by electrostatic gates in a 2D (nonmagnetic) layer which is deposited directly on top of the FM layer. The confinement is assumed to be harmonic in both directions with different confinement lengths $l_x \gg a_x$ and $l_y\gtrsim a_y$ (see Fig.~\ref{fig:system}). The QD under consideration is in the single-particle filling regime, with the lowest orbital level occupied. An orbital level splitting $\gtrsim 10\,$meV is assumed.

In order to couple resonantly with the chiral magnon the qubit splitting is required to be close enough in energy to the edge states in the topological gap, approximately at an energy $E/(JS) = 1.2$ for zero momentum in Fig.~\ref{fig:spectrum}(a). This is ensured by the strong exchange interaction emerging between the FM and the excess electron occupying the QD. This can be achieved if the conduction band edge (hosting the QD) is close enough to the conduction band of the FM allowing for tunnelling, and consequently for exchange interaction between the QD spin and the  spins of the FM lattice. Even though the qubit experiences the large exchange field of the FM layer, $J^\perp \sim 1\,$meV,  the spectrum of the magnet remains unaffected because the nonmagnetic qubit layer remains unpolarized and the QD spin has only a small weight on the individual lattice sites. Here, the contribution of the dipole field is neglected since it is assumed to be sufficiently small, $\Delta_\text{dip}<1\, \mu$eV (see App.~\ref{app:FMQD}) when compared to the exchange field.

Taking the interlayer exchange interaction into account as an effective Zeeman field, the corresponding qubit Hamiltonian reads
\begin{align}
H_\text{SQ} =- J^\perp \sum_i |\psi_\text{QD} (x_i,y_i)|^2 \vec S_i \cdot \boldsymbol \sigma \approx - w J^\perp S \sigma^z \equiv \, \Delta \sigma^z\, ,
\label{eq:HSQ}
\end{align}
where $J^\perp$ is the interlayer exchange interaction strength (i.e., between the FM and the QD layer), $\psi_\text{QD}$ is the orbital part of the QD wavefunction, and $\vec{\sigma}$ is the spin vector-operator with $\sigma^z = \tfrac 1 2 (\ket{\uparrow}\bra{\uparrow} - \ket{\downarrow}\bra{\downarrow})$ being the $z$ component of the QD spin. Furthermore, we used the fact that in the ground state of the FM  $\vec S_i = S \vec e_z$, and therefore the weight of the QD $w = \sum_i |\psi_\text{QD} (x_i,y_i)|^2 \leq 1 $ can be factored out. The localized spin on the QD can be identified with a qubit with basis states $\ket 0 \equiv \ket \uparrow$ and $\ket 1 \equiv \ket \downarrow$ and a qubit splitting $\Delta$.  

The spins of the FM point in the positive $z$ direction: $\langle \vec S \rangle_{T=0} = S \vec e_z$. Therefore, if the ground state $\ket{\downarrow}$ of the qubit is antialigned with the spins of the FM, for example, due to antiferromagnetic interlayer exchange, $J^\perp < 0$, the splitting $\Delta$ is positive. This property is crucial in order to mitigate magnon-induced relaxation from the higher energy qubit state because the transition $\ket{\uparrow} \rightarrow \ket{\downarrow}$ requires a double spin flip, $S_i^-\sigma^-$ (where $S^- \propto a^\dagger$). This process cannot be assisted by the strong interlayer exchange but only by dipole-dipole interaction that is orders of magnitude weaker.

For qubit applications, it is essential to have means to control the qubit and to have long enough coherence times, simultaneously. If the spin and orbital degrees of freedom are coupled in the QD (i.e., via spin orbit interaction or magnetic field gradient), coherent flipping of the qubit can be realized by electric-dipole-induced spin resonance (EDSR) \cite{golovach2006electric,nowack2007coherent,nadj-perge2010spin-orbit,schroer2011field}. In the present setup, besides intrinsic spin-orbit interaction, the induced dipole-field near the edge of the FM can be leveraged for this purpose \footnote{Here, we estimate the dipole-field-induced Rabi frequency to be tens of megahertz. In the case of EDSR the  Rabi frequency is given by $\nu_\text{Rabi} = \hbar^{-1} E_\text{SO} (eE_y l_y) \Delta \Delta^{-2}_\text{orb}$
, if the driving field is applied in the $y$ direction. In order to estimate the dipole-interaction-induced Rabi frequency, we used $E_y = 0.5\,$V/$\mu$m for the amplitude of the drive and $E_\text{SO} = 0.2\,\mu$eV, that is the maximal coupling (as a function of $d_y$ for our set of parameters) that the inhomogeneous dipole-field $B_\text{eff}^y$ can induce between harmonic oscillator basis states in the $y$ direction.}. This mechanism opens a channel for relaxation as well, via coupling to charge noise and phonons. Nonetheless, due to the weakness of the dipole-dipole interaction we do not expect this to be a severe limitation. 

We note that an additional dephasing mechanism appears near the edge of the FM due to the strong exchange field. Since the exchange field is zero outside the FM, the effective qubit splitting $\Delta (d_y) = -w (d_y)J^\perp S $ depends on the QD position as $w (d_y)  = [1+ \erf (d_y/l_y)]/2$, assuming harmonic confinement for the QD, centred around $y = d_y$. This sharp dependence on exchange coupling would make the qubit extremely vulnerable against fluctuations of $d_y$, e.g., due to charge noise. In the following we assume that this dephasing mechanism is prevented by the device design, an assumption that we return to in Sec.~\ref{sec:discussion}, and focus on the dynamical (i.e., magnon-induced) contributions of the decoherence rates.

Even though some of the requirements above might seem stringent at first, due to generality of the results to be presented, we believe that there is a large range of materials that are compatible with the criteria above and can be stacked on top of each other. We return to a discussion of materials in Sec.~\ref{sec:discussion}. For the numerical results in this work the parameters listed in Tab.~\ref{tab:QD} were used, unless  specified otherwise.

\begin{table}[!b]
\caption{Characteristic parameters of the QD  used in numerical evaluations.}
\label{tab:QD}
\begin{tabular}{l c c}
\toprule
Parameter & Symbol & Numerical value \\ 
\cline{1-3}
Qubit g-factor & $g_\text{\tiny{QD}}$ & $2$ \\ 
QD length along the edge & $2l_x$ & $20\,$nm \\ 
QD length perpendicular to the edge & $2l_y$ & $2\,$nm \\ 
\botrule
\end{tabular} 
\end{table}

\subsection{Coupling to the Ferromagnet}
\label{sec:FMQDcoup}

\begin{table}
\caption{Characteristic parameters of the FM-QD coupling used in numerical evaluations.}
\label{tab:coupling}
\begin{tabular}{l c c}
\toprule
Parameter & Symbol & Numerical value \\ 
\cline{1-3}
Interlayer exchange interaction & $J^\perp$ & $-1.2\,$meV
 \\ 
QD-QD distance & $d$ & $1\, \mu$m \\ 
QD distance from the FM edge & $d_y$ & $-0.4\,$nm \\ 
Interlayer distance & $d_z$ & $0.7\,$nm \\ 
\botrule
\end{tabular} 
\end{table}

Assuming a general, non-local coupling $\vec{\hat V}_\text{int} (\vec r_i - \vec r)$ between spin $\vec{S}_i$ and QD spin at position $\vec{r}$, the interaction Hamiltonian between a  qubit and the FM spins can be written as $V_p = \sum_{i} \vec S_{i} \cdot \sum_{\vec r}\vec{\hat V}_\text{int} (\vec r_i - \vec r) \vec{\sigma}_p(\vec r)$, where $\vec{\sigma}_p(\vec{r})$ is the spin density of the $p$th QD and the interaction matrix $\vec{\hat V}_\text{int} (\vec r_i - \vec r)$ contains both exchange and dipolar interactions. As long as SOI is negligible in the QD, the spin $\boldsymbol \sigma$ of the particle in the QD is independent of the spatial coordinates and we may make the ansatz
\begin{align}
   \vec{\sigma}_p (\vec r) = |\psi_p(\vec r)|^2 \boldsymbol \sigma_p\, ,
    \label{eq:ansatzQDspin}
\end{align}
where $\boldsymbol \sigma_p$ is acting on the spin space of the $p$th QD. The spatial part of the QD wavefunction is $\psi_p(\vec r) = \psi(\vec r - \vec r_p)$ is localized around $\vec r_p = (x_p,d_y,d_z)$ and we assume that the two QD wavefunctions have no common support. Thus, we can introduce the coupling matrix between the $p$th QD spin and the $i$th FM spin as $\vec{\hat M} (\vec r_p - \vec r_i) = \sum_{\vec r}\vec{\hat V}_\text{int} (\vec r_i - \vec r)|\psi(\vec r-\vec r_p)|^2$. For notational convenience we introduce the coupling vector $\vec{M}^\alpha = (M^{\alpha x},M^{\alpha y},M^{\alpha z})$ that is the $\alpha$th row of the coupling matrix $\vec{\hat M}$, and $\vec{M}^\pm = \vec{M}^x \pm i \vec{M}^y$, and similarly $\sigma^\pm = \sigma^x \pm i \sigma^y$.

Writing the convolution between the FM spins and the coupling matrix $\vec{\hat M}$ in Fourier space and expanding the coupling terms to first order in magnon creation operators, one obtains
\begin{align}
\begin{split}
    V_p = &\, \mu_B S \vec B_\text{eff}\cdot \boldsymbol \sigma_p \\ 
    &+\frac 1 2 \sum_{k_x, n} \left( \mathrm{e}^{\mathrm{i} k_x x_p}  S^+_{-k_x ,n} \vec{M}^-_{k_x,n} \cdot \boldsymbol \sigma_p + h.c. \right) + \mathcal O (S^0)\, ,
\end{split}
\label{eq:qubittoFMcoupling}
\end{align}
where $\mu_B \vec B_\text{eff} \approx-  w J^\perp \vec e_z$ is the effective field of the FM ground state acting on the qubit as in Eq.~\eqref{eq:HSQ}, while second order terms in magnon creation operators are neglected. The coupling vector connecting the eigenmodes of the FM to one of the qubits is $\vec{M}^-_{k_x,n}=\frac 1 {\sqrt{N_x}}\sum_i e^{-i k_x x_i} \varphi_{-k_x, n}^{\mu_i} (y_i) \vec{M}^- (x_i, y_i - d_y)$. Furthermore, owing to the hermiticity of the Hamiltonian, the coupling matrix elements satisfy $M^{+-}_{k_x, n} = (M^{-+}_{-k_x, n})^*$, $M^{++}_{k_x, n} = (M^{--}_{-k_x, n})^*$, and $M^{+z}_{k_x, n} = (M^{-z}_{-k_x, n})^*$, where $M^{-\pm} = M^{-x} \pm i M^{-y}$.

Now, let two spin qubits (SQs) be situated near the edge of the FM at positions $\vec{r}_\text{\tiny{QD1}} = (-d/2,\,d_y,\,d_z)$ and $\vec{r}_\text{\tiny{QD2}} = (d/2,\,d_y,\,d_z)$, respectively (see Fig.~\ref{fig:system}). The model Hamiltonian under consideration is then
\begin{align}
H = \Delta \left( \sigma_1^z + \sigma_2^z \right) + H_\text{FM}\, + \tilde V ,
\end{align}
where $\Delta = -J^\perp S$, assuming $w=1$ for simplicity, and 
\begin{align}
    \tilde V = \sum_{k_x, n} S^+_{-k_x ,n}  \vec{\ M}^{-}_{k_x,n} \cdot \left( \mathrm{e}^{-\mathrm{i} k_x d/2} \boldsymbol{\sigma}_1 + \mathrm{e}^{\mathrm{i} k_x d/2}  \boldsymbol{\sigma}_2 \right)+h.c.
\end{align}
is the coupling between the two qubits and the magnon modes of the FM. Finally, the parameters of the FM-QD coupling used in our numerical results are listed in Tab.~\ref{tab:coupling}, unless otherwise specified.

\subsection{Effective qubit-qubit coupling}
\label{sec:effQDQDcoup}

In this section we calculate the effective qubit-qubit coupling mediated by the ferromagnet. To this end, we integrate out the magnons from the Hamiltonian by means of a second order Schrieffer-Wolff transformation, and write the effective Hamiltonian as
\begin{align}
H_\text{eff} = \Delta \left( \sigma_1^z + \sigma_2^z \right) + W_\text{eff}\, .
\label{eq:qubitandFM}
\end{align}
The effective coupling between the qubits assumes the form \cite{Bravyi_2011, trifunovic2013long}
\begin{align}
    W_\text{eff} = -\frac{\mathrm{i}}{2\hbar} \lim \limits_{\eta \rightarrow 0^+} \int \limits_0^\infty dt\, \mathrm{e}^{-\eta t} \left\langle \left[\tilde V(t), \tilde V \right] \right\rangle_\text{FM} \, ,
\label{eq:Weff}
\end{align}
where $\eta$ is the lifetime of the intermediate virtual excitation (i.e., magnons). The expectation value $\left\langle \cdots \right\rangle_\text{FM}$ is taken with the FM ground state $\ket{0}_\text{FM}$, and $\tilde V(t) = \mathrm{e}^{\mathrm{i} H_0 t} \tilde V \mathrm{e}^{- \mathrm{i} H_0 t}$ with $H_0 = H_\text{FM} + \Delta ( \sigma_1^z + \sigma_2^z )$. 

Within the framework of the Schrieffer-Wolff transformation it is possible to (implicitly) account for the fact that the pure magnons are not the true eigenstates of the FM. Magnons are dressed by other quasiparticles, e.g., by phonons, causing a finite spectral width of the magnon modes. Equation~\eqref{eq:Weff} can be written in the frequency domain as
\begin{align}
    W_\text{eff} = \frac 1 {2 \hbar} \int \limits_{-\infty}^\infty \frac{d\omega}{2\pi}  \frac{\left\langle \left[ \tilde V(\omega), \tilde V \right] \right\rangle_\text{FM}}{\omega + \mathrm{i} \eta} \, ,
\label{eq:Weffo}
\end{align}
where $\tilde V(\omega) = \int_{\text{\tiny{$-\infty$}}}^{\text{\tiny{$\infty$}}}\! dt\, \tilde V(t) \mathrm{e}^{-\mathrm{i}\omega t}$ and $\hbar \eta$ is the linewidth broadening of the corresponding magnon. In our case the linewidth broadening of the magnon mode $(k_x,n)$ is associated with Gilbert damping and therefore $\hbar \eta\rightarrow \aG \varepsilon_{k_x,n}$, which effectively smears out the magnon density of states cutting unphysical singularities. See App.~\ref{app:SWdiss} for more technical arguments.

Expanding $H_0$ on the eigenbasis of the FM and the qubits, the time evolution of spin and qubit creation operators takes the form $S^-_{-k_x,n}(t) \sigma^-(t) = \sqrt{2S} \mathrm{e}^{\mathrm{i}\varepsilon_{k_x,n}t/\hbar - \mathrm{i}\Delta t/\hbar} a_{k_x,n}^\dagger \sigma^-$. The Fourier transform $\tilde V(\omega)$ then contains terms like $2\pi\hbar \sqrt{2S}M^{++}_{k_x,n} \delta(\hbar \omega-\varepsilon_{k_x,n} + \Delta) a_{k_x,n}^\dagger \sigma^-$, facilitating the exact evaluation of the integral in Eq.~\eqref{eq:Weffo}. Performing the expectation value over magnons, the resulting qubit-qubit interaction can be written as $W_\text{eff} = \sum_{p,q \in \{1,2\}}W_{pq}$, where $W_{pq}$ contains products of qubit operators $\sigma_p$ and $\sigma_q$. Expressing each contribution in terms of the susceptibility in Eq.~\eqref{eq:chinmkx}, we obtain for $p \ne q$ 
\begin{align}
\begin{split}
    W_{pq} = &\frac 1 {32} \sum_{k_x,n} \mathrm{e}^{\mathrm{i}k_x (x_p-x_q)} \chi_{nn}^\perp(\Delta/\hbar, k_x) M^{++}_{k_x,n} \sigma_p^- \\
    &\times\left( M^{-+}_{-k_x,n} \sigma_q^- + M^{--}_{-k_x,n} \sigma_q^+ + M^{-z}_{-k_x,n} \sigma_q^z \right) + \text{h.c.}\, ,
\label{eq:W12}
\end{split}
\end{align}
where we have dropped the off-resonant terms proportional to $\chi^\perp_{nm}(0,k_x)$ and $\chi^\perp_{nm}(-\Delta/\hbar,k_x)$ because they are highly suppressed for antiferromagnetic interlayer exchange $J^\perp < 0$  in the relevant limit, $d \gg l_x \gg a_x$ (see App.~\ref{app:effcoup} for further details). The diagonal terms $W_{pp}$ simply give a tiny dynamical contribution $\delta \vec B_\text{eff}$ to the effective exchange field $\vec B_\text{eff}$. Since $|\delta \vec B_\text{eff}| \ll |\vec B_\text{eff}|$, we omit $W_{pp}$ in $W_\text{eff}$.

The fact that the coupling term of the XY type (i.e., $\propto \sigma^-_1 \sigma^+_2$) is proportional to $|M^{++}|^2$ instead of $|M^{-+}|^2$ is a direct consequence of the antiferromagnetic coupling ($\Delta \propto -J^\perp > 0$). Furthermore, we note that the dipole-dipole interaction can contribute to all terms in Eq.~\eqref{eq:W12}, while the isotropic direct exchange only contributes to the $M^{-+}$ matrix element. The characteristic energies of these two interactions are strikingly different: for the dipole-dipole interaction $\tfrac{\mu_0 \mu_B^2}{a^3} \sim 0.6\,\mu$eV gives an upper bound, while the direct exchange coupling is $|J^\perp| \sim 1\,$meV.  Thus, the strongest coupling term is expected to be $\propto \sigma^-_1\sigma^-_2$. The full analytical form of the coupling matrix elements for the exchange and the dipole mechanisms will be shown below in Secs.~\ref{sec:exanal}-\ref{sec:dipanal}.

\subsection{Decoherence rates}
\label{sec:decoh}

In the previous section the (virtual) magnon-mediated effective qubit-qubit interaction has been discussed. However, the coupling of the QD spin to the ferromagnet also gives rise to decoherence of the spin qubits caused by real magnons. In order to calculate the contribution of magnons to the decoherence times, we decompose the FM-QD interaction Hamiltonian of Eq.~\eqref{eq:qubittoFMcoupling} such that $V = \tfrac 1 2 (V^+ \sigma^- + V^- \sigma^+) + V^z \sigma^z$ and define the corresponding noise power spectra as $\NoSp_{V^b} (\omega)= \int\! dt \Big \{ [ V^b (t) ]^\dagger, V^b (0)\Big \}\, \mathrm{e}^{-\mathrm{i}\omega t}$. The relaxation and dephasing times within the Bloch-Redfield approximation then read as $\Gamma_1 = \tfrac 1 { 4 \hbar^2}\NoSp_{V^-} (\Delta/\hbar)$ and $\Gamma_2^* = \tfrac 1 {4 \hbar^2}\NoSp_{V^z} (0)$, respectively \cite{trifunovic2013long,makhlin2003dissipation}. 

Substituting in the corresponding couplings, to lowest order of the $1/S$ expansion, we can relate both the longitudinal, $\NoSp_{V^z} (\omega)$, and transversal, $\NoSp_{V^-} (\omega)$,  noise spectrum to the transversal magnonic power spectrum $\NoSp^\perp_{k_x,n}(\omega) = \hbar \coth (\beta \hbar \omega /2) \text{Im} [\chi^\perp_{nn}(\hbar \omega, k_x)]$, {where $\beta = (k_\text{B} T)^{-1}$ with $T$ being the temperature. Finally, for the decoherence rates, one obtains
\begin{subequations}
\label{eq:G12BR}
\begin{align}
    \Gamma_1 &= \frac 1 {16 \hbar^2} \sum \limits_{k_x,n} |M^{++}_{k_x,n}|^2 \NoSp^\perp_{k_x,n}(\Delta/\hbar)\, ,\label{eq:relaxation} \\
    \Gamma^*_2 &=  \frac 1 {2 \hbar^2} \sum \limits_{k_x,n} |M^{+z}_{k_x,n}|^2 \NoSp^\perp_{k_x,n}(0) + \mathcal O(S^0).
    \label{eq:dephaserate}
\end{align}
\end{subequations}
The dephasing rate $\Gamma^*_2$ can be highly suppressed when the ferromagnetic resonance is shifted to finite energies, for example, by an external magnetic field or an easy-axis anisotropy ($\Delta_\text{F} > 0$, cf.~Sec.~\ref{sec:FM}).

The appearance of $|M^{++}_{k_x,n}|^2$ in the formula for the relaxation rate of Eq.~\eqref{eq:relaxation} can be understood as follows: for antiferromagnetic coupling ($\Delta \propto -J^\perp > 0$) if no magnons are excited, the excited state of the qubit is $\ket{0_\text{m}\! \uparrow}$ which can then relax to the qubit ground state $\ket{1_\text{m}\! \downarrow}$ creating a magnon by means of the coupling $S^-M^{++}\sigma^-$ (note that $S^- \propto a^\dagger$), where we used the simplified notation $\ket{0_\text{m}}$ for the FM ground state and $\ket{1_\text{m}}$ for a single magnon excitation with energy $\Delta$. For the ferromagnetic case ($\Delta \propto -J^\perp < 0$) the qubit states are reversed and the transition $\ket{0_\text{m}\! \downarrow} \rightarrow \ket{1_\text{m}\! \uparrow}$ describes the relaxation requiring an interaction term of the type $S^-M^{+-}\sigma^+$. The relaxation mechanism for ferromagnetic interlayer coupling is then mediated by direct exchange interaction, as opposed to dipole-dipole interaction in the antiferromagnetic case. For a detailed derivation, we refer the reader to App.~\ref{app:decoh}.

One of the central figure of merits in the field of quantum computing is the gate fidelity $\mathcal F = 2\text{Tr} [\rho (t_\text{op}) \rho_f]-1 \in [0,1] $ that describes the deviation of the qubit state after the operation from the targeted final state, quantified by the respective density matrices $\rho (t_\text{op})$ and $\rho_f$ \cite{poyatos1997complete,loss1998quantum}. Let us consider a two-qubit gate implemented by the time evolution under the static Hamiltonian $W_\text{eff}$. Neglecting the subleading two-qubit terms and the single qubit terms in the time evolution for simplicity, one may consider only the $\sigma^-_1 \sigma^-_2$ coupling to get $\mathrm{e}^{-\mathrm{i} W_\text{eff} t_\text{op} / \hbar} \ket{00} = (\ket{00} -\mathrm{i} \mathrm{e}^{\mathrm{i}\phi}\ket{11})/\sqrt 2$ where $t_\text{op} = \hbar \tfrac \pi 4  |\!\bra{11}W_\text{eff}\ket{00} \!|^{-1}$ is the operation time and $\phi = \arg (\bra{11}W_\text{eff}\ket{00})$. This two-qubit operation, supplemented with single-qubit rotations, i.e., $U_{\!\sqrt{\text{\tiny{SWAP}}}} \sim \sigma^x_1 \mathrm{e}^{-\mathrm{i} W_\text{eff} t_\text{op} / \hbar} \sigma^x_1$, is equivalent to a $\sqrt{\text{SWAP}}$ gate up to a phase. Since the relaxation rate $\Gamma_1$ describes the decay of the diagonal elements of the density matrix as $(\rho_{00}-\rho_{11})(t) \propto \mathrm{e}^{-\Gamma_1 t}$, using the operation time in the exponent, the fidelity of the two-qubit gate is obtained as
\begin{equation}
    \mathcal F =\exp \left(- \frac \pi 4 \frac{\hbar \Gamma_1}{ |\!\bra{11}W_\text{eff}\ket{00} \!|} \right)\sim 1- \frac \pi 4 \frac{\hbar \Gamma_1}{ |\!\bra{11}W_\text{eff}\ket{00} \!|}\, ,
\label{eq:infidelity}
\end{equation}
provided that the decoherence is primarily caused by relaxation. As it will be shown later, this is indeed the case for the chiral mode due to the resonant coupling.

\section{Results}

\begin{figure*}
\includegraphics[width=\textwidth]{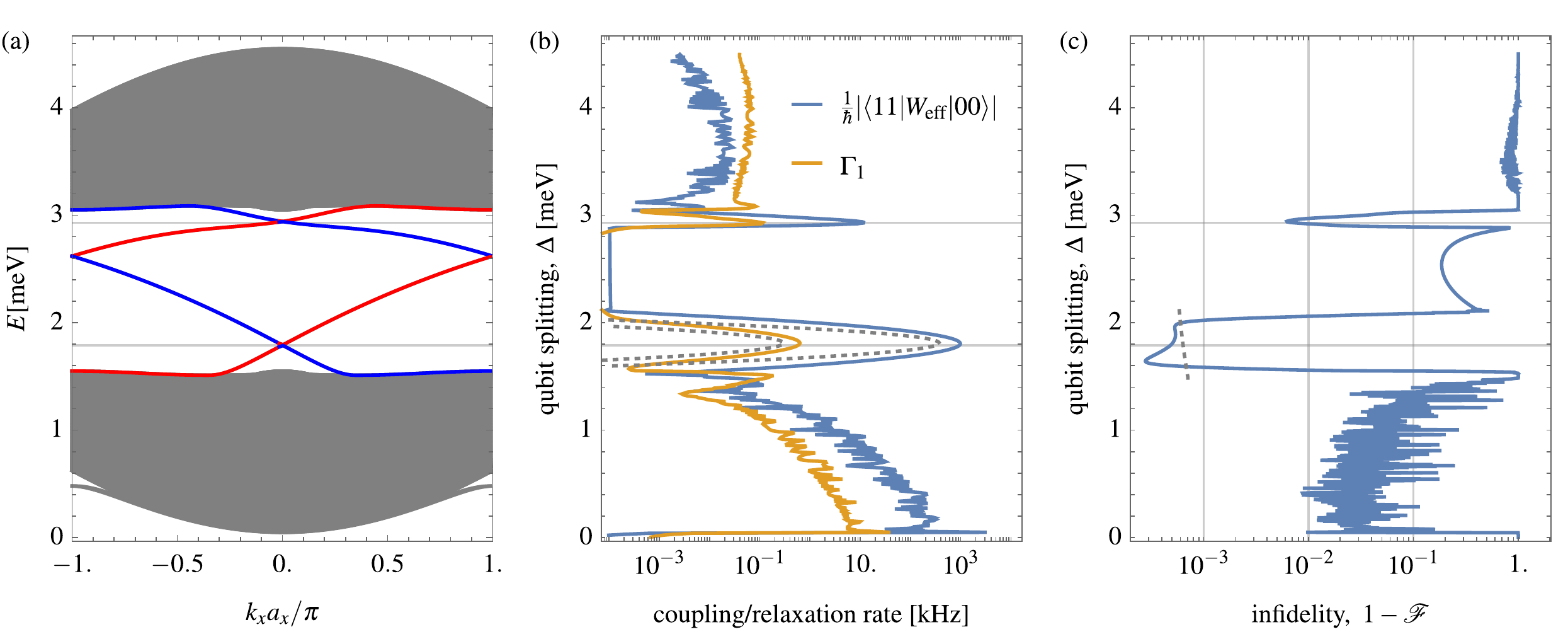}
\caption{(a) Spectrum of the ferromagnetic slab with $N_y = 1000$ unit cells as a function of momentum $k_x$. Chiral edge magnons at opposite edges are indicated by red and blue lines, respectively. (b) Effective qubit-qubit coupling strength (blue) and magnon-induced qubit-relaxation rate (orange) for $T= 100\,$mK, as a function of qubit splitting. Pronounced in-gap resonances are identified and associated with chiral edge magnons. Note that at the two resonances the coupling strength largely exceeds the relaxation rate, making them the optimal operation points for two-qubit gates. (c) Infidelity $1-\mathcal F$ as a function of qubit splitting. Largest fidelities are found in the energy windows of the chiral edge magnons. Vertical lines correspond to a fidelity of $90\%,$ $99\%$, and $99.9\%$, respectively. In (a)-(c), horizontal gray lines indicate the energy of the chiral edge modes at $k_x = 0$. Dashed grey lines in (b) and (c) correspond to the result of the analytical formulas, i.e. Eqs.~\eqref{eq:allW12coupschiral}-\eqref{eq:G1chiral} with the coupling matrices taken from Eqs.~\eqref{eq:MkxJ}~and~\eqref{eq:dipdip}.}
\label{fig:coup_fid}
\end{figure*}

A numerical simulation of the effective coupling has been performed by evaluating $W_{pq}$ in Eq.~\eqref{eq:W12}. To this end we solved the eigenvalue equation of Eq.~\eqref{eq:HFMkx} numerically using a slab unit cell consisting of $N_y=1000$ armchair cells (i.e., $4000$ spins). The resulting eigenvalues $\varepsilon_{k_x, n}$ were used to obtain the susceptibility according to Eq.~\eqref{eq:chinmkx}, while the eigenmodes $\vec \varphi_{k_x,n}$ were used in the explicit expressions for the direct exchange and dipole-dipole couplings [for which we refer the reader to Eqs.~\eqref{appeq:Mex}~and~\eqref{appeq:Mkxndip} of the corresponding appendices]. The interlayer exchange was varied together with the qubit splitting as $J^\perp = -\Delta/S$ to account for the effective exchange field, and the other parameters are listed in Tabs.~\ref{tab:FM}-\ref{tab:coupling}. 

We obtained pronounced Gaussian resonances for both the coupling strength and the relaxation rate when the qubit splitting matches the energy of the edge modes at $k_x=0$ [see Fig.~\ref{fig:coup_fid}(a)-(b)]. 
The coupling strength for the lower in-gap resonance can reach up to $1\,$MHz facilitating fast two-qubit operations over $\mu$m distances with fidelities exceeding $99.9\%$ [see Fig.~\ref{fig:coup_fid}(c)]. Further insight into the dependence of the coupling on the various parameters of our model can be obtained via the analytical formulas within the continuum approximation to be presented below. First we obtain the coupling strength as a function of the coupling matrices in Sec.~\ref{sec:chiral}, then provide analytical formulas for the coupling matrices for both the exchange and dipole-dipole interaction in Secs.~\ref{sec:exanal}-\ref{sec:dipanal}.

\subsection{Chiral edge mode}
\label{sec:chiral}

If the qubit splitting $\Delta$ lies within the magnonic gap and is close to $\varepsilon_0 \equiv \varepsilon_{0,\mathrm{e}}$, defined as the energy of the chiral edge mode at $k_x = 0$, the effective coupling in Eq.~\eqref{eq:W12} simplifies as the contribution of bulk modes ($n\neq \mathrm{e}$) are far off-resonant. Since the spin density of the QD is distributed over several lattice sites, the qubit spin $\sigma$ can only couple to magnon modes with $k_x \lesssim l_x^{-1}$, where the spectrum of the edge mode can be written as $\varepsilon^{}_{k_x,\mathrm{e}} = v_x k_x + \varepsilon_0 $, with $v_x \sim 0.39\,$meV$\cdot$nm.
Finally, the susceptibility near the edge resonance reads as
\begin{align}
    \chi_{nn}^\perp(\Delta, k_x) \approx -2S \frac{\delta_{n \mathrm{e}}}{v_x (k_x + \mathrm{i} \kappa) - \delta}
\end{align}
where $\delta = \Delta - \varepsilon_0$ is the detuning from the edge resonance and $\kappa^{-1} \approx  \frac{v_x}{\aG \varepsilon_0}$ is the mean free path of the chiral magnon. In the continuum approximation we replace the sum over $k_x$ by an integral [as in Eq.~\eqref{appeq:Wijintegral}]. Furthermore, close to resonance the integration limit can be extended to infinity, provided that $|\delta/v_x| \ll \pi/a_x$. Then, exploiting that $\vec{\hat M}^{}_{k_x,n}$ is an analytic function of $k_x$, we can perform the momentum integral using the residue theorem [see Eq.~\eqref{appeq:Wijresidue}] to obtain
\begin{subequations}
\label{eq:allW12coupschiral}
\begin{align}
    \bra{01} W_\text{eff} \ket{10}
\approx & \,  -\mathrm{i} \frac{S a_x}{16v_x}  \mathrm{e}^{\mathrm{i} k_0 d-|\kappa| d} |M^{++}_{k_0,\mathrm{e}}|^2 \, \label{eq:coupXY} ,\\
    \bra{11}  W_\text{eff} \ket{00} 
 \approx &\, \mathrm{i}\frac{S a_x}{16v_x} \mathrm{e}^{-\mathrm{i} |k_0| d-|\kappa| d} M^{++}_{k_0,\mathrm{e}} M^{-+}_{-k_0,\mathrm{e}}\, ,\\
    \bra{10}  W_\text{eff} \ket{00} 
\approx & \, \mathrm{i} \Theta (v_x)\frac{S a_x}{16|v_x|} \mathrm{e}^{-\mathrm{i}|k_0| d-|\kappa|d} M^{++}_{k_0,\mathrm{e}} M^{-z}_{-k_0,\mathrm{e}}\, ,\\
    \bra{01} W_\text{eff} \ket{00}
\approx & \, \mathrm{i} \Theta (-v_x)\frac{S a_x}{16|v_x|} \mathrm{e}^{-\mathrm{i}|k_0| d-|\kappa| d} M^{++}_{k_0,\mathrm{e}} M^{-z}_{-k_0,\mathrm{e}}\, ,
\end{align}
\end{subequations}
where $k_0 = \delta/v_x$ and we neglected $\kappa$ in the coupling, i.e., $M^{+\alpha}_{k_0+\mathrm{i}\kappa,\mathrm{e}} \approx M^{+\alpha}_{k_0,\mathrm{e}} + \mathcal O (\kappa d_y)$. This latter approximation is justified since every length scale in the coupling is much smaller than $\kappa^{-1} \approx 2.2\, \mu \text m$, for example, $d_z, d_y \sim 1\,$nm.

The Gaussian dependence on the detuning around the resonance in Fig.~\ref{fig:coup_fid}(b) can be understood via the spatial averaging effect of the QD. Since the magnetic moment of the particle is equally distributed along the QD, the coupling to magnon modes with $k_x > l_x^{-1}$ is averaged out leading to $\vec{\hat M}^{}_{k_x,n} \propto \mathrm{e}^{-k_x^2l_x^{2}/4}$ (see App.~\ref{app:ex}). Furthermore, $l_x$ is much larger than the remaining length scales in the coupling (i.e., $l_y$, $d_y$, $d_z$, and $a$) and therefore one can expand the coupling as $\vec{\hat M}^{}_{k_x,n} \approx \mathrm{e}^{-k_x^2l_x^{2}/4} \vec{\hat M}^{}_{0,n} + \mathcal{O} (l_y/l_x)$.

Using the same assumptions as for the effective coupling, the contribution of the edge modes to the decoherence rates can also be estimated using Eq.~\eqref{eq:G12BR}. If the detuning is close to zero, the relaxation is dominated by the resonant edge mode at $k_x = 0$ and reads as 
\begin{align}
    \Gamma_1 \approx  \frac{S a_x}{16 \hbar v_x} |M^{++}_{k_0,\mathrm{e}}|^2 \coth (\beta \varepsilon_0/2)\, .\label{eq:G1chiral}
\end{align}
Since the (pure) dephasing rate is proportional to $\NoSp^\perp_{k_x,n} (0)$ [see Eq.~\eqref{eq:dephaserate}], the contribution of the edge mode is far off-resonant. In order to estimate it, we expanded the susceptibility in the Gilbert damping $\aG$ to get $\NoSp^\perp_{k_x,e} (0) \approx \hbar \coth(\beta \varepsilon_0/2) 2\aG S /\varepsilon_0$. The dephasing rate can then be written as
\begin{align}
    \Gamma^*_{2,\mathrm{e}} \approx \frac{\aG S}{\hbar \sqrt{2 \pi} \varepsilon_0} \frac{a_x}{l_x} |M^{-z}_{0,\mathrm{e}}|^2 \coth (\beta \varepsilon_0/2)\, , \label{eq:G2chiral}
\end{align}
where we exploited that $M_{k_x,n} \approx \mathrm{e}^{-k_x^2l_x^{2}/4} M_{0,n}$ for every mode. Using Eq.~\eqref{eq:G2chiral} we obtain $\Gamma^*_2 \sim 10^{-4}\,$Hz that is vastly underestimating the dephasing rate (as it will be shown later).

In order to find the leading contribution to dephasing we need to consider the modes that are closest to zero energy. We do this in the 2D limit, which is valid deep in the bulk when the QD is far from the edges of the FM. Here one can replace the coupling $M_{k_x,n}$ by $M_{k_x,k_y}$, that is the coupling to the magnon mode with energy $\varepsilon_{k_x, k_y}$, as obtained for periodic boundary conditions along the $x$ and the $y$ direction. Since $l_x\gg a_x$, we still restrict ourselves to $k_x = 0$ in the coupling to get
\begin{align}
\Gamma^*_2 &\approx \frac{\aG S}{\hbar \sqrt{2 \pi} \Delta_F} \frac{a_x}{l_x}  \coth (\beta \Delta_F/2) \sum_{k_y} |M^{-z}_{0,k_y}|^2\, ,
\label{eq:G2FMR}
\end{align}
where we neglected the curvature of the magnon band since $\Delta_F\lesssim \varepsilon_{0,k_y}$. 

Using Eq.~\eqref{eq:coupXY} and Eq.~\eqref{eq:G1chiral}, an important relation can be deduced, namely,
\begin{align}
    \frac{|\bra{01}  W_\text{eff} \ket{10}|}{\hbar \Gamma_1} = \mathrm{e}^{-|\kappa| d} \tanh (\beta \varepsilon_0/2) \sim \mathcal O (1)\, ,
    \label{eq:XYperRel}
\end{align}
meaning that the relaxation provides an upper bound for the XY coupling, regardless of the strength of the QD-FM coupling. The same formula is valid in the ferromagnetic interlayer coupling regime ($J^\perp > 0$), where both quantities are proportional to $|M^{+-}|^2$. Since the $\bra{01}  W_\text{eff} \ket{10}$ is the leading coupling in that case, virtual magnon processes are unable to create entanglement between qubits while maintaining the coherence of the two-qubit system. Therefore, we have focused here on the antiferromagnetic case, where the strongest coupling is $\bra{11}  W_\text{eff} \ket{00}$; we will revisit the ferromagnetic coupling case in Sec.~\ref{sec:transd}, where we try to leverage the fast magnon emission/absorption rate ($\Gamma_1$) in a scheme where a real magnon is mediating the coupling between distant spin qubits (as opposed to virtual magnons considered so far).

The dependence of the coupling on the inter-QD distance $d$ is explicitly defined in Eq.~\eqref{eq:allW12coupschiral}, however, in order to determine the coupling strength and to identify the dependence on the QD size and position we need to calculate the coupling matrix elements $M_{k_x,e}$ for the case of direct exchange and dipole-dipole coupling.

\subsection{Exchange coupling}
\label{sec:exanal}

\begin{figure}
    \includegraphics[width= \columnwidth]{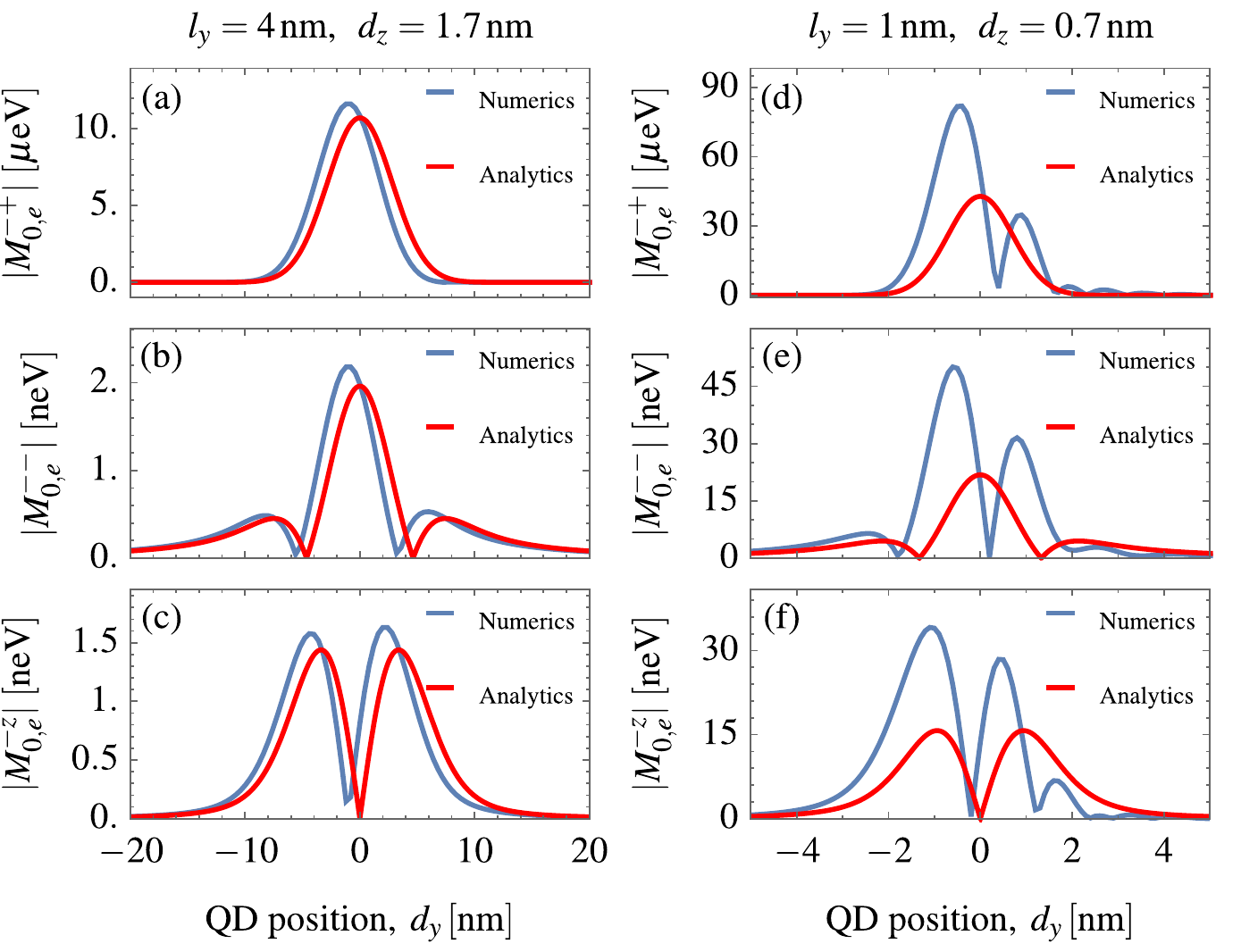}
    \caption{Coupling matrix elements $M^{-+}_{k_x,\mathrm{e}}$, $M^{--}_{k_x,e}$, and $M^{-z}_{k_x,e}$ at $k_x = 0$ including both direct exchange and dipole-dipole interactions as a function of the QD position $d_y$ for (a)-(c) $l_y = 4\,$nm and $d_z=1.7\,$nm and (d)-(f) $l_y = 1\,$nm and $d_z=0.7\,$nm. 
    Numerical results (blue lines) are obtained from Eqs.~\eqref{appeq:Mex}~and~\eqref{appeq:Mkxndip}. Analytical results (red lines) are given in  Eqs.~\eqref{eq:MkxJ}~and~\eqref{eq:dipdip}. The rest of the parameters are given in Tabs.~\ref{tab:FM}-\ref{tab:coupling}. Direct exchange interaction only contributes to $M^{-+}$, therefore this coupling element is orders of magnitude larger than $M^{--}$ and $M^{-z}$. We observe good quantitative agreement between the numerical and analytical curves in (a)-(c), whereas for smaller $l_y$ and $d_z$ values in (d)-(f) the coupling is strongly asymmetric around the FM edge ($d_y = 0$) due to the spatial profile of the edge magnon that is not taken into account in the analytics.}
    \label{fig:Mkx_dy}
\end{figure}

Let us first consider the contribution of (isotropic) direct exchange interaction between the FM spin $\vec S_i$ and the qubit spin $\boldsymbol \sigma$ that is given by the exchange matrix $-\mathbf{\hat J}_i |\psi (\vec r_i -  \vec r_\text{\tiny{QD}})|^2$, where $\mathbf{\hat J}_i$ is the local spin-spin interaction matrix between the $i$th site of the FM and the qubit layer. In this case the effective coupling between the magnonic mode $(k_x,n)$ and the spin qubit is given by
\begin{align}
\begin{split}
    M^{-+}_{-k_x,n} =& -\frac 1 {\sqrt{N_x}}\sum_{\vec r_i,\mu} \mathrm{e}^{\mathrm{i} k_x (x_i-x_\text{\tiny{QD}})}  \varphi^{\mu}_{k_x, n} (y_i) 2 J^\perp_i |\psi (\vec r_i\!+\!\vec r^\mu -\! \vec r_\text{\tiny{QD}})|^2 \\
\approx& -\mathrm{e}^{-k_x^2 l_x^2/4}  J^\perp C_{k_x,n}\, ,
\end{split}
\label{eq:MkxJ}
\end{align}
where we assumed the QD wavefunction to be Gaussian, i.e., $|\psi (\vec r_i)|^2 = \tfrac{a_x}{\sqrt \pi l_x} \mathrm{e}^{-x_i^2/l_x^2} |\psi (y_i)|^2$, and we defined $C_{k_x,n} = \tfrac{1}{2} \sum_{y_i,\mu} \varphi^{\mu}_{k_x, n} (y_i) \mathrm{e}^{\mathrm{i} k_x x^\mu} |\psi (y_i - d_y)|^2$. Furthermore, for simplicity we assumed  homogeneous and isotropic coupling, $\mathbf{\hat J}_i \approx J^\perp\, \mathds{1}$. 

Owing to the Gaussian factor in the coupling the main contribution of the coupling matrix to the qubit-qubit coupling in Eq.~\eqref{eq:allW12coupschiral} is given by small momenta ($k_x \lesssim l_x^{-1}$). In this regime, $C_{k_x,\mathrm{e}} \approx C_{0,\mathrm{e}}$ is a good approximation for the coefficient in Eq.~\eqref{eq:MkxJ}. Provided that the DMI is strong enough, i.e., $D>0.1J$, the localization length of the edge mode is small, i.e., $\lambda \ll l_y$ [see Fig.~\ref{fig:spectrum}(c)]. In this limit we can factor out $\delta \mu_\text{e}$ and estimate the corresponding coefficient as $C_{0,\mathrm{e}} \approx \tfrac{a_y}{2\sqrt \pi l_y} \delta \mu_{\mathrm{e}} \mathrm{e}^{-d_y^2/l_y^2}$ (see App.~\ref{app:ex} for further details).

\begin{figure*}
\includegraphics[width=\textwidth]{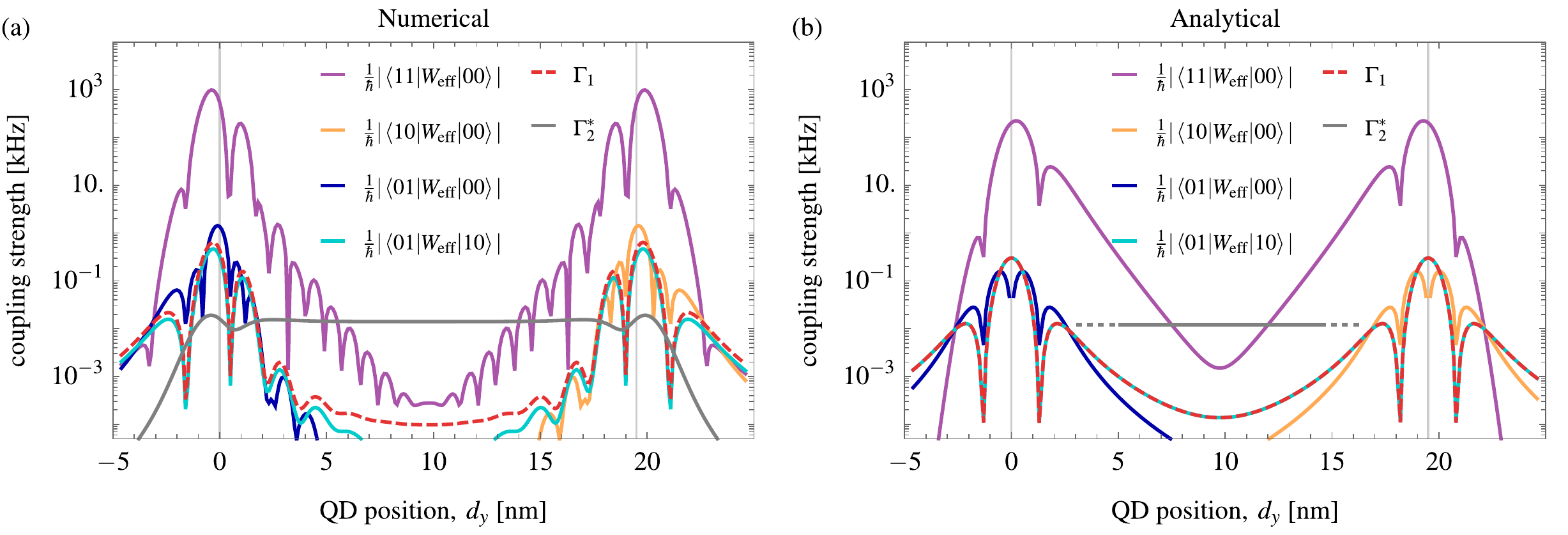}
    \caption{
    Position dependence of the effective qubit-qubit couplings and the qubit relaxation. The parameters are the same as for Fig.~\ref{fig:coup_fid}, except for the slab width  that is set to $N_y = 20$ in order to show both edges of the FM simultaneously. (a) Numerical results for the matrix elements of the effective coupling $W_\text{eff}$ of Eq.~\eqref{eq:W12} for a constant qubit splitting of $\Delta = 1.8\,$meV. (b) Analytical results of the couplings and the relaxation are given in Eqs.~\eqref{eq:allW12coupschiral}-\eqref{eq:G1chiral} by substitution of Eqs.~\eqref{eq:MkxJ} and \eqref{eq:dipdip}. The analytical estimate in the bulk for $\Gamma_2^*$ was calculated using Eq.~\eqref{eq:G2FMR} with Eq.~\eqref{eq:intMkx0ky}. The excellent quantitative agreement between the numerical simulation (a) and analytical formulas (b) facilitates the estimation of the various coupling and decoherence time scales in different materials without having to perform the heavy numerical calculations.}
\label{fig:coup_decoh_dy}
\end{figure*}

\subsection{Dipole-dipole coupling}
\label{sec:dipanal}
Owing to its long-ranged nature, calculations involving the dipole-dipole interaction are unwieldy and deferred to App.~\ref{app:dip}. Here, we only note that the exponential suppression factor in momentum,  $\mathrm{e}^{-k_x^2 l_x^2/4}$, appears regardless of the form of the interaction potential. Therefore, we restrict our attention to the $k_x = 0$ case and provide an analytical formula, assuming that the other confinement length of the QD, $l_y$ is sufficiently large, compared to the localization length of the edge magnon, $\lambda$, e.g., $\lambda \ll l_y$. The coupling matrix elements in this limit read as 
\begin{subequations}
\label{eq:dipdip}
\begin{align}
    M_{0,\mathrm{e}}^{-+} = -M_{0,\mathrm{e}}^{--} =&\frac{\mu_0 }{\pi} \frac{g\, g_\text{\tiny{QD}} \mu_B^2}{a_x l_y^2}  \delta \mu_{\mathrm{e}} \text{Re} 
    \left[
     I \left( \tfrac{\mathrm{i} d_y-d_z}{l_y} \right)
     \right] ,
\end{align}
\begin{align}
    M_{0,\mathrm{e}}^{-z} =& -\mathrm{i} \frac{\mu_0 }{\pi} \frac{g\, g_\text{\tiny{QD}} \mu_B^2}{a_x l_y^2} \delta \mu_{\mathrm{e}} \text{Im} \left[ I\left( \tfrac{\mathrm{i} d_y-d_z}{l_y} \right)\right],
\end{align}
\end{subequations}
where we have introduced the complex function $I(x) = 1 + \sqrt \pi  x \mathrm{e}^{x^2} [1+\text{erf}(x)]$. Similarly to the case of the direct exchange interaction, it is the $\lambda \ll l_y$ assumption that allowed us to factor out the dynamical magnetic moment $\delta \mu_\mathrm{e}$ of the chiral edge mode from the integral.

The analytical estimates for the couplings $M^{-+}_{0,e}$, $M^{--}_{0,e}$, and $M^{-z}_{0,e}$ obtained in Eqs.~\eqref{eq:MkxJ}~and~\eqref{eq:dipdip} are compared with the numerically evaluated exact expressions given in Eqs.~\eqref{appeq:Mex} and \eqref{appeq:Mkxndip} as a function of $d_y$ in Fig.~\ref{fig:Mkx_dy}. The analytical formulas are in very good agreement with the numerics as shown in Fig.~\ref{fig:Mkx_dy}(a)-(c) for $l_y = 4\,$nm and $d_y=1.7\,$nm. Further parameters of the QD and the FM were set as in Tabs.~\ref{tab:FM}-\ref{tab:coupling}. The only apparent deviation is the slight shift of the peaks in the numerics, compared to the edge ($d_y = 0$). We attribute this effect to the asymmetric nature of the edge mode (i.e., the mode terminates with a sharp maximum at the edge and decays exponentially towards the bulk) that is not taken into account in the analytical estimate which assumes $\varphi^\mu_{0,\mathrm{e}} (y) \sim \tfrac{1}{4} \delta\mu_\mathrm{e} \delta (y)$. 

The FM-QD coupling matrix elements are presented in Figs.~\ref{fig:Mkx_dy}(d)-(f) for the same parameters used in Fig.~\ref{fig:coup_fid}, i.e., $l_y = 1\,$nm and $d_z = 0.7\,$nm. Even though the localization length $\lambda \sim 1\,$nm, is comparable with $l_y$, the qualitative behavior is correct and the maximal coupling strength is reliable in order of magnitude. As compared to the analytical prediction, the numerical results exhibit features that are slightly shifted outwards from the edge [similarly to Fig.~\ref{fig:Mkx_dy}(a)-(c)] and small oscillations appear on the side of the FM ($d_y>0$). These effects appear due to the spatial ``fine structure'' of the edge mode (exponential decay and oscillations on the scale of $a_y$) that is not accounted for in the analytical approximation. 

In order to complement the estimate of the bulk dephasing formula in Eq.~\eqref{eq:G2FMR}, we provide here the relevant coupling matrix element for $k_x = 0$ as a function of $k_y$ (assuming periodic boundary conditions along $y$ direction). The coupling between the QD spin and the lower-energy acoustic magnon band reads 
\begin{align}
M^{-z}_{0,k_y} \approx  \frac{\mu_0 \mu_B^2 g g_\text{\tiny{QD}}}{2 a_x a_y} k_y \mathrm{e}^{-k_y^2l_y^2/4} \mathrm{e}^{-|k_y|d_z}\, ,
\label{eq:Mkx0ky}
\end{align}
where we neglect the contribution of the optical magnon band, since their contribution is suppressed by the negligible dynamical magnetic moment as well as the large energy denominator in Eq.~\eqref{eq:G2FMR}. The sum over $k_y$ modes can be evaluated in the continuum limit  as
\begin{align}
\sum_{k_y} |M^{-z}_{0,k_y}|^2 \approx \frac 1 {2\pi} \frac{a_y}{l_y}  \left( \frac{\mu_0 \mu_B^2 g g_\text{\tiny{QD}}}{2 a_x a_y l_y} \right)^2 I_2(d_z/l_y)\, ,
\label{eq:intMkx0ky}
\end{align}
where $I_2 (x) = \sqrt{2\pi} (1+4x^2)\mathrm{e}^{2x^2} [1+\erf (\sqrt 2 x)] - 4x $.

\subsection{Position dependence of the effective coupling}
\label{sec:dydep}

Choosing a smaller system size of $N_y = 20$ and considering the various couplings and decoherence rates as a function of $d_y$ allows us to compare the full analytical formulas with the numerics in Fig.~\ref{fig:coup_decoh_dy}(a)~and~(b). For this we have tuned the qubit energy to be on resonance with the $k_x = 0$ chiral mode i.e., $\Delta = 1.8\,$meV [see Fig.~\ref{fig:coup_fid}(a)]. Even though, in potential experiments if the QD is moved outside the FM (in-situ), the decreasing interlayer exchange experienced by the QD would tune the qubit frequency out of resonance \footnote{Within the present assumptions the detuning from resonance would change as $\delta = \frac {S|J^\perp|} 2 \{\erf [d_y/l_y
] + \erf [(L_y-d_y)/l_y]\} - \varepsilon_0$. Note that depending on the value $J^\perp$, the resonance can be reached at any $d_y$ in principle.}, which is not taken into account in Fig.~\ref{fig:coup_decoh_dy}.

As shown in Fig.~\ref{fig:coup_decoh_dy}, the peaks of the effective coupling develop only close to the two edges of the sample at $d_y = 0$ and $d_y = 20\,$nm, which provides a natural way to tune the qubits in and out of the coupling regime. This property is crucial since the qubit splitting is set by the interlayer exchange interaction that is challenging to tune in-situ. On the other hand $d_y$ can be changed freely in the range $d_y \in [0,20\,\text{nm}]$ since the interlayer exchange is constant to a good approximation in this range.
 
The strongest coupling is achieved for the $\bra{11}  W_\text{eff} \ket{00}$ matrix element because this is the only coupling that is proportional to the interlayer exchange $J^\perp$. In order to capture the exponential decay towards the bulk we have used Eq.~\eqref{appeq:MkxJerf} for the analytical curve instead of the simplistic formula for $C_{0,\mathrm{e}}$ given in Sec.~\ref{sec:exanal}.

The second strongest coupling are the $\sigma^\pm \sigma^z$-type of terms that come about three orders of magnitude smaller than the $\bra{11}  W_\text{eff} \ket{00}$ term. Importantly, since the propagation direction is opposite along the left and right edges, from Eq.~\eqref{eq:allW12coupschiral} we expect only $\bra{01}  W_\text{eff} \ket{00}$ coupling on the left edge (because $v_x < 0$) and $\bra{10}  W_\text{eff} \ket{00}$ on the right edge (because $v_x > 0$). This is fulfilled up to several orders of magnitude in Fig.~\ref{fig:coup_decoh_dy} (cf.~yellow and blue lines) and a clear marker of chirality.
 
The excellent agreement between numerical and analytical results is sustained for the decoherence rates as well. The ratio of the XY coupling and the relaxation rate in Eq.~\eqref{eq:XYperRel} being $\mathcal O (1)$ is confirmed by the numerical results close to the edge resonances. The dephasing rate estimate for the bulk in Eq.~\eqref{eq:G2FMR} [using Eq.~\eqref{eq:intMkx0ky}] turns out to be a very good estimate a few nm away from the edges.

\section{Qubit entanglement via chiral magnon transduction}
\label{sec:transd}

As mentioned in Sec.~\ref{sec:chiral}, the virtual magnon coupling strength and the relaxation rate are both proportional to the coupling $|M^{+-}_{0,e}|^2$ in the ferromagnetic interlayer coupling regime ($J^\perp>0$). Therefore, since $|\bra{01} W_\text{eff} \ket{10}| \sim \Gamma_1$, the virtual magnon mediated coupling is inefficient in this case. One possibility to overcome this limitation is through coupling by {\it real} magnons (as opposed to virtual ones described above). Provided that the FM-QD coupling (i.e., $J^\perp$) can be switched on and off on demand, the first qubit can be used to emit a chiral edge magnon that propagates and is subsequently absorbed by the second qubit, coherently. This protocol is leveraging that the emitted magnon wave packet will propagate towards the second qubit maintaining its shape (quasi-linear dispersion) because it cannot backscatter at defects due to its chirality and the presence of the topological gap. 

Previous proposals for such a magnon transduction protocol have focused on the single magnon mode approximation \cite{fukami2021opportunities}. Such an approximation, however, can only be made if the energy separation from higher magnonic modes is much larger than the coupling strength. For the case of the chiral magnon this energy scale is $v_x/\mathcal C \sim 1\,$neV, where $\mathcal C \sim 100\,\mu$m is a typical circumference of the sample and $v_x \approx 0.39\,$meV$\cdot$nm. This energy separation is orders of magnitude smaller than the achievable FM-QD coupling $g\propto M_{0,e}^{+-} $ in Fig.~\ref{fig:Mkx_dy}(a). Note that we use Fig.~\ref{fig:Mkx_dy}(a) as a reference here, because the coupling is dominated by direct exchange interaction $J^\perp$ and therefore it agrees up to a sign with the coupling of the ferromagnetic case ($J^\perp>0$).

In order to discuss the limit where $g \gg v_x/\mathcal C$, the complete dynamics of the local magnon excitations need to be considered. To model the scenario when the spin qubits are on resonance with the chiral edge mode, we consider a one-dimensional bosonic lattice with the dispersion relation given by the chiral edge mode. In order to reduce the computational cost further, we extend the FM unit cell to several lattice sites, $a_x\rightarrow 2l_x$, thereby backfolding the spectrum as depicted in Fig.~\ref{fig:transduct}(a). We consider only a single mode that used to cross $k_x = 0$ before the backfolding of the spectrum, which is the red line in the highlighted area in Fig.~\ref{fig:transduct}(a). The coupling to higher-energy edge modes (originally at $k_x = n \pi/l_x$) is negligible, since they are suppressed by the factor $\mathrm{e}^{-(n \pi)^2/4}$. Furthermore, we assume that each spin qubit couples to a single FM unit cell and the coupling is uniform within the unit cell. 

In order to mitigate the contribution of the virtual magnon processes, only one of the qubits should be coupled to the magnon mode at any given time. The entangling protocol then consists of three steps, viz., (i) first qubit is coupled and emits a magnon; (ii) both qubits are decoupled and the magnon propagates; (iii) second qubit is coupled and absorbs the magnon. This could be achieved, for example, when the effective coupling strength is $g = 5\,\mu$eV, in which case the size of the emitted magnon wave packet $v_x \hbar/g \sim 80\,$nm is indeed much smaller than the distance between the two qubits.

\begin{figure}
 \includegraphics[width=1.\columnwidth]{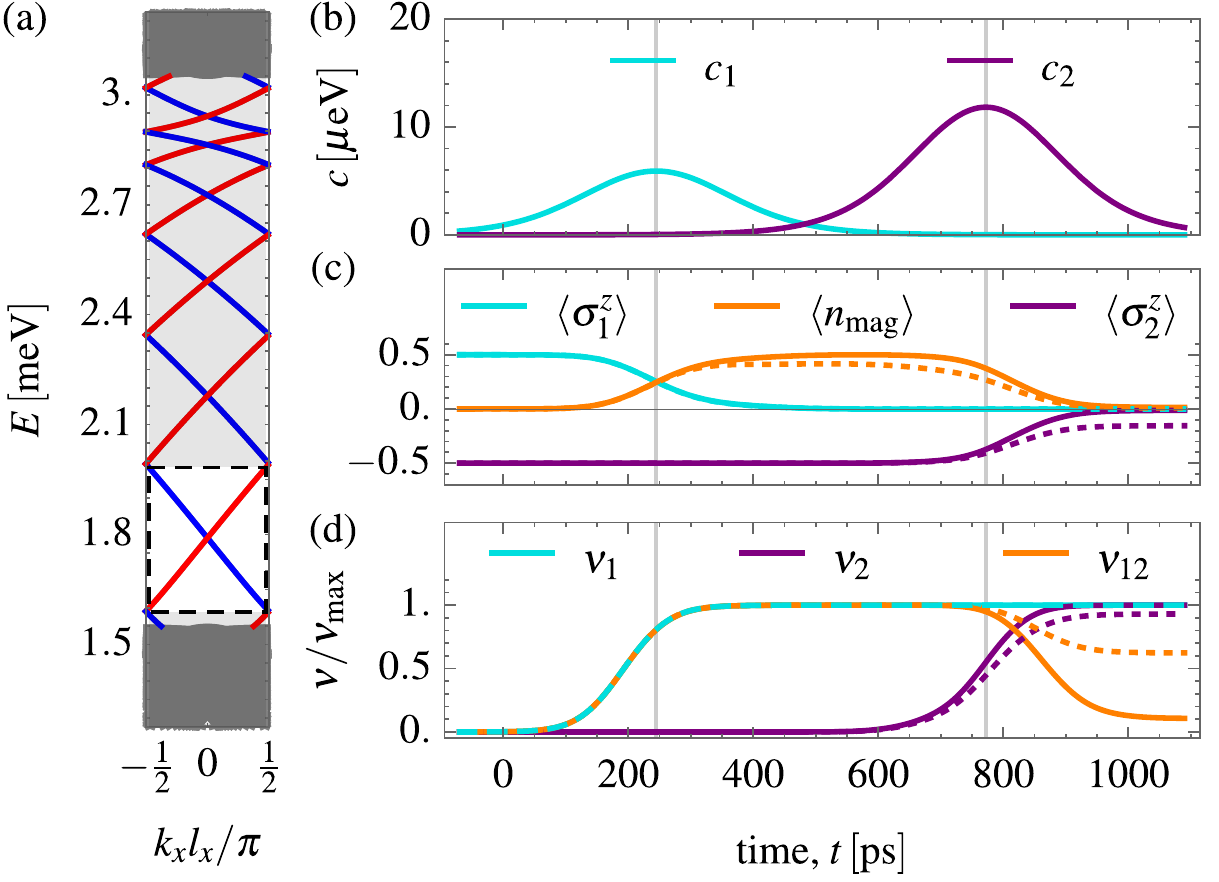}
\caption{(a) Backfolded topological magnonic bandgap of the ferromagnet with a unit cell size of $2l_x = 5 a_x$. Only the coupling to the highlighted bands are relevant (dashed white box), as discussed in the main text. (b) Coupling strength $c_{1,2}$ as a function of time for the first and second qubit, respectively. (c) Expectation value of the qubit spins $\sigma_1^z$ and $\sigma_2^z$ and the total magnon number $n_\text{mag}$ in the edge of the ferromagnet. (d) von Neumann entropy of the first qubit $\nu_1$, the second qubit $\nu_2$, and the two-qubit system $\nu_{12}$. Solid lines correspond to the dissipationless process and dotted lines to $\aG = 10^{-4}$ and $\Gamma_2^* = 100\, $kHz. It is apparent from (c) and (d) that even small dissipation has a detrimental effect on the entanglement of the qubits with the environment (i.e., $\nu_{12} \neq 0$ in the end of the protocol) suggesting that efficient preparation of a two-qubit entangled state requires a magnon mean-free-path that is much longer than the qubit-to-qubit distance.}
\label{fig:transduct}
\end{figure}

We account for the local coupling and the dynamics of the emitted magnon wave packet by performing a numerical simulation of the system by solving the Lindblad equation~\cite{lindblad1976generators} 
\begin{align}
\begin{split}
    \dot \rho =& - \tfrac{\mathrm{i}}{\hbar} [ H(t), \rho ] +\tfrac{1}{2} \Gamma_2^* \sum \limits_{i=\{1,2\}} D[\sigma_i^z] \rho\\
&+ \frac{2\aG}{\hbar}\sum \limits_{k} \varepsilon_k \left\{ (1+\bar n )\mathcal D[a^{}_k] \rho +  \bar n \mathcal D[a^\dagger_k] \rho \right\}
\end{split}
\label{eq:lindblad}
\end{align}
where $ \mathcal D[O] \rho = O \rho\, O^\dagger - \tfrac 1 2 ( O O^\dagger \rho + \rho\, O O^\dagger )$ and $\bar n = [\exp (\beta\varepsilon_k) -1]^{-1}$, with $\varepsilon_k$ being the linear dispersion of the edge mode. The corresponding Hamiltonian is written as
\begin{align}
\begin{split}
    H(t) =& \Delta(\sigma_1^z+\sigma_2^z)+ c_1(t,\delta t)a_1\sigma^+_1 \\
&+ c_2(t- t_\text{prop},\delta t) a_{N_x} \sigma^+_2 + \text{h.c.},
\end{split}
\end{align}
where $t_\text{prop} = \hbar d/v_x$ is the propagation time. The time dependence of the coupling is a smeared out box function, i.e., $c_i(t,\delta t) = g_i n_F [t/t_\text{rise}] n_F [(t+\delta t)/t_\text{rise}]$, where $\delta t$ is the length of the pulse, $n_F (x) = (\exp (x) +1)^{-1}$ and $t_\text{rise} = 70\,$ps is the rise time \footnote{The choice of the rise time is a crucial step in order to create a magnon wave packet that can be efficiently absorbed by the second qubit. We found that a pulse with (relatively) long $t_\text{rise}$ time creates a more symmetric wave packet that can be absorbed with a higher accuracy (i.e., magnon number reduces close to zero after absorption).} of the pulse [see Fig.~\ref{fig:transduct}(b)]. 

From the time series of the density matrix, we evaluate the von-Neumann entropy of the $i$th qubit, defined by 
\begin{equation}
\nu_i = -\underset{\small{ \mathcal H \backslash \mathcal H_i}}{\text{Tr}}[\rho \ln(\rho)]\, ,
\label{eq:entanglemet}
\end{equation}
where $\underset{\small{ \mathcal H \backslash \mathcal H_i}}{\text{Tr}}$ denotes the partial trace, excluding the subspace of the corresponding qubit. Additionally, $\nu_{12}$ and  $\nu_\text{m}$ are the entropies of the two-qubit system and the magnons with the environment, respectively. The spin expectation value of the $i$th qubit is then calculated as
\begin{equation}
\langle \sigma_i  \rangle = \underset{\small{ \mathcal H \backslash \mathcal H_i}}{\text{Tr}}[\rho\, \sigma_i] \, ,
\label{eq:spinexp}
\end{equation}
and the magnon occupation number is
\begin{equation}
\langle n_\text{mag} \rangle = \underset{\small{ \mathcal H \backslash \mathcal H_m}}{\text{Tr}}[\rho\, \Sigma^{}_l a^\dagger_l a^{}_l] \, .
\label{eq:nmag}
\end{equation}

The time-evolution of the entenglement entropy $\nu_i$ of Eq.~\eqref{eq:entanglemet}, the spin expectation value $\langle \sigma_i \rangle$ of Eq.~\eqref{eq:spinexp}, and the magnon number $\langle n_\text{mag} \rangle$ of Eq.~\eqref{eq:nmag}  are presented in Fig.~\ref{fig:transduct}(c)-(d). The density matrix $\rho (t)$ in the definition of these quantities were obtained by numerically integrating Eq.~\eqref{eq:lindblad}. We set $2l_x = 14 a_x$ and consider $N_x = 11$ lattice sites along the one-dimensional chain. After the emission of the magnon, the wave packet propagates along the chain twice, traveling a total distance of $42l_x \approx 510\,$nm before it is reabsorbed by the second qubit (periodic boundary conditions have been assumed). Two cases are differentiated: dissipationless [$\aG =0$ and $\Gamma^*_2 = 0$; see solid lines in Fig.~\ref{fig:transduct}(c)-(d)] and dissipative evolution [$\aG =10^{-4}$ and $\Gamma^*_2 = 100\, $kHz; see dotted lines in Fig.~\ref{fig:transduct}(c)-(d)]

The dissipationless case can be discussed straightforwardly in the state vector representation. At $t=0$, the time evolution starts from a pure state of each subsystems i.e.,
$\ket{\uparrow}_1 \ket{\downarrow}_2 \ket{0}_m$. Between $t = 200\,$ps and $t = 300\,$ps the first qubit emits a magnon with $50\%$ probability leading to an entangled state $ \tfrac{1}{\sqrt{2}}[\ket{\uparrow}_1 \ket{\downarrow}_2 \ket{0}_\text{m} + \ket{\downarrow}_1 \ket{\downarrow}_2 \ket{\psi(x)}_\text{m}]$ where $\ket{\psi(x)}_\text{m}$ is a spatially extended wavepacket of a single mangon. At this point the first qubit has a vanishing spin expectation value and the magnon occupation number is $1/2$. The entanglement is created between the first qubit and the magnon, therefore $\nu_{1} = \nu_\text{max}\equiv \ln 2$ and $\nu_{12} = \nu_\text{m} = \nu_\text{max}$ [see solid lines in Fig.~\ref{fig:transduct}(c)-(d)]. Until $t = 700\,$ps the magnon wave packet propagates through the lattice and reaches the position of the second qubit ($\ket{\psi(x)}_\text{m} \rightarrow \ket{\psi(x-d)}_\text{m}$; \footnote{Since the first qubit is coupled to the first unit cell and the second to the last unit cell, the effective qubit-qubit distance in the simulation is $d = 42 l_x \sim 510\,$nm.}). In the final step of the protocol the second qubit needs to absorb the incoming magnon with $100\%$ probability [thus the doubled coupling strength in Fig.~\ref{fig:transduct}(b)] creating a pure state of the two qubit system
$\tfrac{1}{\sqrt{2}}[\ket{\uparrow}_1 \ket{\downarrow}_2 + \ket{\downarrow}_1 \ket{\uparrow}_2] \ket{0}_\text{m}$ with $\nu_{1} = \nu_{2} = \nu_\text{max}$ and $\nu_{12} = \nu_\text{m} = 0$.

When the qubit decoherence and the Gilbert damping in Eq.~\eqref{eq:lindblad} is included, the main difference compared to the dissipationless case is the damping of the magnon wavepacket during its propagation, i.e., $\langle n_\text{mag}\rangle < 1/2$ at $t = 700\,$ps in  Fig.~\ref{fig:transduct}(c). The entanglement with the environment can be tracked via the entanglement entropies in Fig.~\ref{fig:transduct}(d). Since the magnon number goes to zero, the corresponding entanglement $\nu_\text{m}$ needs to vanish (the vacuum of magnons is a pure state). Nontheless $\nu_{12} \neq 0$, meaning that the two qubits are still entangled with another subsystem, the environment. Therefore, considering $\nu_{1} = \nu_\text{max}$ or $\nu_{2} \approx \nu_\text{max}$ gives a false impression about $\nu_{12}/\nu_\text{max} \approx 0.6$ becomes the only appropriate measure for the infidelity.

We conclude this section by noting that even though qubit-qubit entanglement can be created through magnon transduction, the fidelity of the operation is seriously limited by the magnon mean free path (as also pointed out in Ref.~\cite{fukami2021opportunities}). Moreover, we note that such a resonant coupling protocol does not correspond to universal two-qubit logic~\cite{blais2021curcuit} with the chirality of the edge magnon restricting the quantum computing applications of the magnon transduction protocol even further.

\section{Discussion}
\label{sec:discussion}

In the setup considered in this Paper, the QD needs to be close to the FM edge in order to achieve sufficiently strong coupling. However, at this position, due to the large exchange field gradient, the qubit is susceptible to fluctuations of its position (i.e., $\delta d_y$), which would lead to a fluctuating qubit splitting and thus dephasing. In order for the dephasing rate to stay well below the two-qubit operation frequencies, the fluctuations $\delta d_y$, need to be small enough; we estimate that $\mathrm{e}^{-d_y^2/l_y^2} (\langle \delta d_y^2\rangle /l_y^2)^{-1/2} \ll 10^{-6}$.  There are different ways to overcome this limitation: (i) The QD can be confined in a narrow nanowire along the FM edge that fixes its position and therefore the effective exchange field. In that case, it is required that the qubit splitting is tuneable by other means (e.g., via an external field if $g_\text{\tiny{QD}} \neq g_\text{\tiny{FM}}$) and thereby the qubit-qubit coupling can be switched on and off on demand [as in Fig.~\ref{fig:coup_fid}(b)]. (ii) The qubit can be located close to a domain-wall in DMI instead of the edge, where the magnetization is constant throughout but the DMI strength $D$ changes sign. Since the chirality of the edge mode is given by $\text{sign}(D)$, for a given ground state magnetization~\cite{Owerre2016a}, two well-localized edge modes are propagating in the same direction along such a domain wall, potentially increasing the coupling strength by a factor of two. (iii) The QD layer can be terminated as well at the edge of the FM layer. In this case the QD experiences a constant interlayer exchange $J^\perp$, and therefore the decoupling has to be performed by moving the QD towards the bulk of the lattice. Additionally, option (i) and (iii) might offer a solution to achieve a QD that is narrow enough $l_y \sim 1\,$nm to efficiently couple to the edge mode.

Throughout this work we concentrated on a honeycomb-lattice topological magnon insulator. This model is approximately realized in monolayers of the van der Waals materials CrI$_3$ \cite{Chen2018CrI3}, CrSiTe$_3$, and CrGeTe$_3$ \cite{Zhu2021}. These materials support chiral edge magnons in the low THz range. The honeycomb-lattice model is also realized in artificial arrays of magnetic disks hosting topological magnetic solitons that interact magnetostatically; chiral modes are found in the low GHz range \cite{Kim2017Soliton}. The general formulas for the effective two-qubit coupling derived here, for example, in Eq.~\eqref{eq:allW12coupschiral}, are agnostic to the actual realization of the platform hosting topological magnons. As such, they apply to any topological-magnon host---be it on the honeycomb or other lattices, in the GHz or THz range---and provide a guide for the identification of suitable materials. 

Finally, we point out that the long-range spin-qubit entanglement may also be used as a probe for the experimental detection of topological chiral edge magnons, one of the key challenges in the field of topological magnons \cite{Malki_2020, Li_2021, McClarty_2021, Wang_2021}. Chiral magnetic edge excitations are notoriously hard to detect with common probes of magnetism that are nonlocal and mostly bulk-sensitive, such as inelastic neutron scattering. In contrast, the \emph{local} coupling to quantum-dot spin qubits, as shown above, can be considerably large. A single spin qubit probes the local magnonic density of states via relaxation processes. By taking the difference of detuning-resolved relaxation times at the edges with that in the bulk, one can verify the existence of edge-located in-gap magnon modes. Moreover, the detuning dependence of the relaxation time is remarkably distinct for linearly dispersing bands, as expected for a chiral mode, and trivial parabolic bands (see Appendix \ref{app:effcoup}). On top of that, the two-qubit setup, in particular the transduction protocol in Sec.~\ref{sec:transd}, provides a direct experimental handle on chirality because the entanglement protocol is unidirectional.

\section{Conclusion}
\label{sec:conclusion}
We have presented a two-layer setup where the FM bottom layer hosts a chiral magnon mode with energy lying in the magnonic band gap. Coupling spin qubits to the chiral magnon facilitates two different long range qubit-qubit coupling protocols, both of which have beeen studied in detail.

Two-qubit coupling can be mediated by virtual magnons. We found that this protocol is efficient if the interlayer exchange interaction $J^\perp$ is antiferromagnetic. For $1\,\mu$m qubit separation, $1\,$MHz coupling strength has been found with a $\sqrt{\text{SWAP}}$ gate fidelity up to $99.9\%$. We also presented general analytical formulas for coupling of two-dimensional spin qubits with chiral edge magnons that can be of great use trying to identify the optimal materials and dimensions for such a system.

Finally, we have investigated the magnon transduction protocol in the ferromagnetic interlayer coupling regime. The coupling is highly fast ($\sim 1\,$GHz), owing to the excitation of a physical magnon. Even though the mean-free path of the edge magnon seriously limits the fidelity of such a two-qubit coupling, the transduction protocol can be used as an experimental probe of the chirality of topological edge magnons.

\begin{acknowledgements}
We thank Stefano Bosco for useful discussions. This work was supported by the Georg H. Endress Foundation and the Swiss National Science Foundation NCCR SPIN. 
\end{acknowledgements}

\appendix

\section{Conventions}
\label{app:conventions}

In this appendix we guide the reader through the conventions we used throughout the main text and give explicit formulas as examples. First we define the Fourier transformation of an operator $O = \sum_i O_i$ as follows
\begin{align}
    O_k \equiv \frac{1}{\sqrt{N}} \sum_{i=1}^N \mathrm{e}^{ - \mathrm{i} k r_i } O_i.
\end{align}
For the bosonic creation and annihilation operators, this convention results in
\begin{subequations}
\begin{align}
    a_k &\equiv \frac{1}{\sqrt{N}} \sum_{i=1}^N \mathrm{e}^{ - \mathrm{i} k r_i } a_i,
    \\
    a_k^\dagger & \equiv \frac{1}{\sqrt{N}} \sum_{i=1}^N \mathrm{e}^{ \mathrm{i} k r_i } a^\dagger_i 
    =  \left(a_k\right)^\dagger,
\end{align}
\end{subequations}
where the corresponding commutation relation is $[a^{}_k,a^\dagger_{k'}] = \delta_{kk'}$. For the Fourier transformation of the FM spin operators in the $x$ direction this leads to
\begin{subequations}
\begin{align}
    S^+_i &\approx \sqrt{\frac{2S}{N_x}} \sum_{k_x} \mathrm{e}^{\mathrm{i} k_x x_i } \sum_{n=1}^{4N_y} \varphi^{\mu_i}_{k_x,n} (y_i) a_{k_x,n}\, ,
    \\
    S^-_i &\approx \sqrt{\frac{2S}{N_x}} \sum_{k_x} \mathrm{e}^{\mathrm{i} k_x x_i } \sum_{n=1}^{4N_y} \left[\varphi^{\mu_i}_{-k_x,n} (y_i)\right]^* a^\dagger_{-k_x,n}\, .
\end{align}
\end{subequations}
where we have performed a transformation from the band index $n$ to the index pair $(y_i,\mu)$ as well, with $y_i$ being the armchair unit cell index, and $\mu$ is the index within the unit cell. Furthermore, the transformed spin operators can be expressed with the Holstein-Primakoff bosons as
\begin{subequations}
\begin{align}
    &S^+_{k_x,n} \approx \sqrt{2S} a_{k_x,n}\, ,
    \\
    &S^-_{k_x,n} \approx \sqrt{2S} a^\dagger_{-k_x,n}\, .
\end{align}
\end{subequations}
Consequently, the time evolution of the transformed spin operators reads as 
\begin{subequations}
\begin{align}
  &  S_{k_x,n}^+(t) \approx e^{-i\varepsilon_{k_x,n}t}  S_{k_x,n}^+\, ,
    \\
  &  S_{k_x,n}^-(t) \approx e^{i\varepsilon_{-k_x,n}t}  S_{k_x,n}^-\, ,
\end{align}
\label{appeq:timeevoSkxn}
\end{subequations}
where we point out that $S_{k_x,n}^+(t) = [S_{-k_x,n}^-(t)]^\dagger$.

We define the susceptibility of the transformed spin operators as
\begin{align}
\begin{split}
    \chi_{nm}^\perp(t,k_x)
    &\equiv - i \theta(t) \delta_{nm} \langle [S_{-k_x,n}^-(t),S_{k_x,n}^+] \rangle \\
    &= i \theta (t) 2S \delta_{nm} e^{i\varepsilon_{k_x,n}t}\, ,
\end{split}
\label{appeq:chitk}
\end{align} 
where we used the time evolution of the spin operators in Eq.~\eqref{appeq:timeevoSkxn}.

Furthermore, in frequency space the susceptibility assumes the form
\begin{align}
\begin{split}
    \chi_{nm}^\perp(\omega,k_x) &= \int_{-\infty}^{\infty}dt\, e^{-i\omega t - \eta t} \chi_{nm}^\perp(t,k_x)\\
    &= \frac{-2S \hbar}{\varepsilon_{k_x,n} - \hbar \omega + i \eta} \delta_{nm}\, ,
\end{split}
\label{appeq:chiok}
\end{align} 
where one can substitute the linewidth as $\eta \rightarrow \aG \varepsilon_{k_x,n}$ in the case of Gilbert damping.

\section{Effective qubit-magnon coupling: analytical formulas}
\label{app:FMQD}

Assuming a general, non-local coupling between the qubit and the ferromagnet spins,  the interaction Hamiltonian can be written as $V_p = \sum_{i} \vec S_{i} \cdot \vec{\hat M} (\vec r_p - \vec r_i) \boldsymbol{\sigma}_p$. Writing the convolution between the FM spins and the coupling matrix $\vec{\hat M}$ in Fourier space, and expanding the coupling terms to first order in magnon creation operators one obtains
\begin{align}
\begin{split}
V_p = &\frac 1 2 \sum_{k_x, n} e^{ik_x x_\text{\tiny{QD}}}  (S^+_{-k_x ,n} \vec{M}^-_{k_x,n}  + S^-_{-k_x ,n} \vec{M}^+_{k_x,n} ) \cdot \boldsymbol{\sigma}_p\\
& + \mu_B S \vec B_\text{eff} \cdot \boldsymbol \sigma_p+ \mathcal O (S^0)\, ,
 \end{split}
\label{appeq:qubittoFMcoupling}
\end{align}
where $\vec B_\text{eff}$ is the effective magnetic field of the FM ground state felt by the qubit. The couplings $\vec{M}^\pm_{k_x,n} = \vec{M}^x_{k_x,n} \pm i \vec{M}^y_{k_x,n}$ have three vector components $x,y,z$ and can be expressed with the real space coupling matrix elements as
\begin{align}
&\vec{M}^-_{k_x,n}=\frac 1 {\sqrt{N_x}}\sum \limits_{x_i,y_i} \sum_\mu e^{-i k_x x_i} \varphi_{-k_x, n}^\mu (y_i) \nonumber\\
& \hspace{70pt} \times\vec{M}^- (x_i + x^\mu, y_i + y^\mu - y_\text{\tiny{QD}}),
\label{appeq:MxytoMkxn}\\
&\vec{M}^+_{k_x,n}=\frac 1 {\sqrt{N_x}}\sum \limits_{x_i,y_i} \sum_\mu e^{-i k_x x_i} \left[\varphi_{k_x, n}^\mu (y_i) \right]^* \nonumber \\
 & \hspace{70pt} \times\vec{M}^+ (x_i + x^\mu, y_i + y^\mu - y_\text{\tiny{QD}}),\\
&\vec{B}_\text{eff}=\sum \limits_{x_i,y_i} \sum_\mu \vec{M}^z (x_i + x^\mu, y_i + y^\mu - y_\text{\tiny{QD}}),
\end{align}
provided that $x_\text{\tiny{QD}}$ is commensurate with the lattice and therefore, the index $x_i$ can be shifted by $x_\text{\tiny{QD}}$.

Assuming that the QD is very narrow, i.e., $l_y^2 \ll d_y^2 +  d_z^2$, the QD is subjected to a homogeneous magnetic field that is given by the dipole field of the FM slab at its position. For the parameters used in the main text $l_y \sim d_y,d_z$, but the approximation above can still be used to estimate the contribution of the dipole-field to the qubit splitting as shown in Fig.~\ref{fig:Mkx_dy}. The dipole-field of the FM ground state may be estimated by that of a magnetized ribbon substituting $\vec m_1(\vec r) = \Theta(y)\Theta(L_y-y) S \vec e_z$ into Eq.~\eqref{appeq:dipcl}, where $\Theta(y)$ is the Heaviside step function. The dipole field felt by the qubit at a position $\vec r = (x, d_y, d_z)$ is then given by
\begin{subequations}
\label{appeq:dipolefield}
\begin{align}
&\mu_B B_\text{eff}^x=0\, ,\\
&\mu_B B_\text{eff}^y =  -\frac{\mu_0 }{4\pi} \frac{z_0g g_\text{\tiny{QD}} \mu_B^2}{a_x a_y} \left[\frac{d_z}{d_y^2+d_z^2}-\frac{d_z}{(L_y\!+\!d_y)^2+d_z^2} \right]\, ,\\
&\mu_B B_\text{eff}^{z} = - J^\perp - \frac{\mu_0 }{4\pi} \frac{z_0 g g_\text{\tiny{QD}} \mu_B^2}{a_x a_y} \nonumber \\ 
&\hspace{75pt}\times\left[-\frac{d_y}{d_y^2+d_z^2}+\frac{L_y\!+\!d_y}{(L_y\!+\!d_y)^2+d_z^2} \right]\, ,
\end{align}
\end{subequations} 
where $L_y = N_y a$, the magnetic moment density of the ribbon is given by $z_0 \mu_B^2 /a_x a_y$ ($z_0=4$ is the number of spins in the FM unit cell), and we included the exchange field as well in the last equation. The contribution of the $z$-component of the dipole field is $\sim 0.6\,\mu$eV for the parameters presented in the main text, and therefore one might neglect it compared to the exchange field. We note that the $g$-tensor anisotropy in the QD layer can be straightforwardly accounted for, by replacing $g_\text{\tiny{QD}} \boldsymbol \sigma$ with $\vec{ \hat g} \boldsymbol \sigma$, where $\vec{ \hat g}$ is the $g$-tensor of the QD.

\section{Including spectral broadening in the Schrieffer-Wolff transformation}
\label{app:SWdiss}

In this appendix we revisit the formula for the second order Schrieffer-Wolff transformation and show how spectral broadening can be included in the subspace to be projected out. For simplicity, in this section we consider a single qubit coupled to the magnons via the effective coupling $V$ of Eq.~\eqref{appeq:qubittoFMcoupling}, but the calculations we provide here can be extended straightforwardly to the two-qubit system. Starting from the definition of $W_\text{eff}$ in the Fourier space given in Eq.~\eqref{eq:Weffo} of the main text, assuming that the linewidth broadening of the qubit is negligible compared to the broadening of the magnons we get
\begin{align}
W_\text{eff} = \frac 1 {2 \hbar } \sum_{\alpha, \beta} \int \limits_{-\infty}^\infty \frac{d\omega}{2\pi} \frac{\bra{0}_\text{FM} \big[ V_{\alpha \beta}(\omega)  \ket{\alpha} \bra{\beta} , V \big] \ket{0}_\text{FM}}{\omega + \mathrm{i} \Gamma [ \varepsilon_{\alpha \beta}(\omega)]} \, ,
\label{appeq:Weffo}
\end{align}
where $\ket{\alpha}, \ket{\beta}$ are qubit basis states corresponding to the energies $\varepsilon_\alpha, \varepsilon_\beta$, respectively. The state $\ket{0}_\text{FM}$ is the vacuum of magnons and $\varepsilon_{\alpha \beta}(\omega) = |\hbar\omega - (\varepsilon_\alpha - \varepsilon_\beta)|$ is the contribution of the magnons to the total excitation energy $\hbar \omega$. Furthermore, the relaxation rate $\Gamma[\varepsilon]$ is the inverse lifetime of the magnon.

Physically the motivation behind this substitution is the following: the magnons are coupled to the phonons of the FM lattice and thereby these modes are dressed. However, the qubits are coupled to each other via the "pure" magnon modes. Therefore, we need to account for the indirect coupling of the two-qubit system to the phonons of the FM lattice through the finite lifetime $\Gamma[\varepsilon]$ of the magnons.

Rewriting the coupling $V$ on the eigenbasis of each subsystems (e.g., the qubits and the ferromagnet) and substituting the corresponding time dependence we get 
\begin{align}
\begin{split}
V_{\alpha \beta}(t)=\sqrt{\frac S 2}\sum \limits_{k_x,n}\  \left[ e^{i\varepsilon_{k_x,n} t/\hbar}a^\dagger_{k_x,n} (\vec{M}^+_{k_x,n} \!\cdot\! \boldsymbol{\sigma})^{}_{\alpha \beta}\right.&\\
\left.+  e^{-i\varepsilon_{k_x,n} t/\hbar}a^{}_{k_x,n}(\vec{M}^-_{-k_x,n}\!\cdot\! \boldsymbol{\sigma})^{}_{\alpha \beta} \right]& e^{i(\varepsilon_\alpha  - \varepsilon_\beta)t/\hbar} .
\end{split}
\end{align}
Taking the Fourier transform in time leads to
\begin{align}
\begin{split}
V_{\alpha \beta}(\omega)= \sqrt{2S} \pi \hbar \sum \limits_{k_x,n}\  \Big[(\vec{M}^+_{k_x,n}\! \cdot\! \boldsymbol{\sigma})^{}_{\alpha \beta} \delta(\Delta_{\alpha \beta}\!+\!\varepsilon_{k_x,n}\!-\!\hbar \omega) a^\dagger_{k_x,n}& \\
 +(\vec{M}^-_{-k_x,n}\! \cdot\!\boldsymbol{\sigma})^{}_{\alpha \beta}\delta(\Delta_{\alpha \beta}\!-\! \varepsilon_{k_x,n}\!-\!\hbar \omega)a^{}_{k_x,n}\Big]& \, ,
\end{split}
\end{align}
with $\Delta_{\alpha \beta} = \varepsilon_\alpha - \varepsilon_\beta$. Substituting $V_{\alpha \beta}(\omega)$ into Eq.~\eqref{appeq:Weffo}, one obtains
\begin{align}
\begin{split}
W_\text{eff} =& \frac S {4 } \sum_{\alpha, \beta}  \ket \alpha \bra \beta \sum \limits_{k_x,n} \sum_{\gamma} (\vec{M}^-_{-k_x,n}\! \cdot\!\boldsymbol{\sigma})^{}_{\alpha \gamma} (\vec{M}^+_{k_x,n}\! \cdot\!\boldsymbol{\sigma})^{}_{\gamma \beta}\\
& \times\left( \frac{1}{\Delta_{\alpha \gamma} - \varepsilon_{k_x,n} + i \hbar \Gamma (\varepsilon_{k_x,n})} + \frac{1}{\Delta_{\beta \gamma} - \varepsilon_{k_x,n} - i \hbar \Gamma (\varepsilon_{k_x,n})}\right) \, ,
\end{split}
\label{appeq:2ndordpert}
\end{align}
that is the usual 2nd order perturbative formula extended with the linewidth broadening of the intermediate state. 

The range of validity can be determined from Eq.~\eqref{appeq:2ndordpert} by requiring that the second order correction $\delta \varepsilon_\alpha =\bra \alpha W_\text{eff} \ket \alpha$ to the qubit energy level $\varepsilon_\alpha$ is much smaller than the orbital level splitting of the QD, assumed to be $\Delta_\text{orb} \sim 10\,$meV, and the bandwidth of the respective magnonic subband $W \propto JS \approx 2\,$meV. In particular, we consider (i) the magnon mode $n$ that is closest to the qubit splitting; (ii) the transition $\alpha = \uparrow$ and $\gamma=\downarrow$, for which $\Delta_{\uparrow \downarrow} = \Delta$. We neglect transitions $\alpha = \gamma$ because there are no resonant transitions for $\Delta_{\alpha \alpha} = 0$ due to the FM resonance gap ($ \varepsilon_{k_x,n} \geq \Delta_F$). The correction to the qubit splitting due the magnon mode $n$ reads as
\begin{align}
\delta \Delta_n = \frac{S a_x}{16\pi} \int\! dk\, |M^{++}_{k_x,n}|^2 \frac{\Delta - \varepsilon_{k_x,n}}{(\Delta - \varepsilon_{k_x,n})^2 + \hbar^2 \Gamma^2 (\varepsilon_{k_x,n})}  \, ,
\end{align}
which we then rewrite in terms of the density of states $\rho_n(\varepsilon) = dk_x/d\varepsilon_{k_x,n}$ as 
\begin{align}
\delta \Delta_n = \frac{S a_x}{16\pi} \int \limits_{\varepsilon_\text{min}}^{\varepsilon_\text{max}}\! d\varepsilon\, \rho_n (\varepsilon)\sum_{\gamma}  |M^{++}_{k_\varepsilon,n}|^2 \frac{\Delta - \varepsilon}{(\Delta - \varepsilon)^2 + \hbar^2 \Gamma^2 (\varepsilon)}  \, ,
\end{align}
where the integration boundaries correspond to the lowest- and highest-energy magnon state of $\varepsilon_{k_x,n}$. We first consider the case, when the qubit splitting is renormalized by a quadratic mode $\varepsilon_{k_x,n} = \varepsilon_0 + D_x k_x^2$ with the density of states $\rho_n(\varepsilon) = [4D_x (\varepsilon-\varepsilon_{0,n})]^{-1/2}$. We exploit that for long QDs the coupling can be estimated as $|M^{++}_{k_x,n}|^2 \approx |M^{++}_{0,n}|^2 e^{-k_x^2l_x^2/2}$. Then, the renormalization of the qubit splitting is given by
\begin{align}
\delta \Delta_n \approx \frac{S a_x}{16\pi} |M^{++}_{0,n}|^2 \int \limits_{0}^{D_xl_x^{-2}}\! d\varepsilon' \frac{1}{2\sqrt{D_x \varepsilon'}} \frac{\delta - \varepsilon'}{(\delta - \varepsilon')^2 + \hbar^2 \Gamma^2}  \, ,
\label{appeq:dDeltak2}
\end{align}
where $\delta = \Delta - \varepsilon_{0,n}$ and we have cut the frequency integral at $D_xl_x^{-2}$ to account for the decay of the coupling $V$ in $k$-space and approximated $\Gamma(\varepsilon)$ with a constant linewidth $\Gamma$. First, we note that for $\Gamma = 0$ the above integral diverges as $1/\sqrt{\delta}$ near resonance. Evaluating the correction in Eq.~\eqref{appeq:dDeltak2} for finite linewidth $\Gamma$ and small detunings $\delta$, we get
\begin{align}
\delta \Delta_n \approx  \frac{S }{32} \frac{|M^{++}_{0,n}|^2}{\sqrt{2D_x a_x^{-2} \hbar \Gamma}} \left[ 1+ \mathcal O (\delta /\hbar \Gamma)\right]  \, ,
\end{align}
where we omitted terms that are $\Gamma l_x^2/ D_x $ and $\delta l_x^2/ D_x$. Importantly, the correction is no longer divergent on resonance, owing to the linewidth that acts as a low-frequency cutoff in this case. For a very conservative estimate we substitute $|M^{++}_{k_x,n}| \sim \tfrac{\mu_0 \mu_B^2}{a^3} \sim 0.6\,\mu$eV and the FM resonance mode with $\Gamma = \aG \Delta_F \sim 5\,$neV that leads to $\delta \Delta_n \lesssim 30\,$neV on resonance, that is well within the $\sim0.5\,$meV bandwidth of the respective magnon mode. Moreover, we note that the density of states is not singular in the 2D limit ($\aG N_y \gg 1$) leading to even larger range of validity for the bulk modes.

Now, we turn to the discussion of the chiral magnon mode $\varepsilon_{k_x,e} = \varepsilon_0 + v_x k_x$ that plays a central role in our work. Following similar considerations as in Eq.~\eqref{appeq:dDeltak2} for the linear mode we get
\begin{align}
\delta \Delta_e \approx  \frac{S a_x}{16\pi v_x} |M^{++}_{0,e}|^2 \int \limits_{-v_x/l_x}^{v_x/l_x}\! d\varepsilon' \frac{\delta - \varepsilon'}{(\delta - \varepsilon')^2 + \hbar^2 \Gamma^2}  \, ,
\label{appeq:dDeltalink}
\end{align}
where we get a finite contribution even for $\Gamma = 0$. In fact for $\delta = 0$, the correction is $\delta \Delta_e = 0$ in Eq.~\eqref{appeq:dDeltalink} because the spectrum is symmetric around the qubit splitting leading to no renormalization of the excited qubit state. However, note that the upper ($\varepsilon>\Delta$) and the lower ($\varepsilon<\Delta$) parts of the integral are both logarithmically divergent if $\Gamma = 0$. Taking the contribution of the magnon modes above resonance ($\varepsilon>\Delta$) into consideration we get
\begin{align}
\begin{split}
\delta \Delta_{e,\varepsilon > \Delta} &\approx  \frac{S a_x}{16\pi v_x} |M^{++}_{0,e}|^2 \int \limits_{0}^{v_x/l_x}\! d\varepsilon' \frac{\delta - \varepsilon'}{(\delta - \varepsilon')^2 + \hbar^2 \Gamma^2} \\
& \approx  - \frac{S a_x}{32\pi v_x} |M^{++}_{0,e}|^2 \left[\log \left( 1 + \frac{v_x^2}{\hbar^2 \Gamma^2 l^2_x}\right)  + \mathcal O (l_x\delta/ v_x)\right] \, ,
\end{split}
\end{align}
where we assume $\delta = 0$ to arrive at the second line. In a very pessimistic estimate we might replace the logarithm with $-2\log{\aG} \approx 18$, which leads to $\delta \Delta_{e,\varepsilon > \Delta} \sim 10^{-2}\,$neV for $|M^{++}_{0,e}| \sim 100\,$neV and the parameters used in the main text. This qubit splitting correction is several orders of magnitude smaller than the bandwidth of the chiral mode.

\section{Effective qubit-qubit coupling}
\label{app:effcoup}

Here, we show first how to obtain Eq.~\eqref{eq:W12} of the main text and the analytical formulas for linear and quadratic magnon modes in the subsequent subsections. To this we use the real time expression for $W_\text{eff}$ defined in Eq.~\eqref{eq:Weff} and the coupling $V$ defined in Eq.~\eqref{appeq:qubittoFMcoupling} in the Heisenberg representation as 
\begin{align}
\begin{split}
\tilde V(t) = \frac 1 2 \sum_{p\in \{1,2\}} \sum_{k_x, n} e^{ik_x x_p} (&e^{i\varepsilon_{k_x,n} t}S^+_{-k_x ,n} \vec{M}^-_{k_x,n} \\
& + e^{-i\varepsilon_{-k_x,n} t}S^-_{-k_x ,n} \vec{M}^+_{k_x,n} )\cdot \boldsymbol{\sigma}_p(t),
 \end{split}
\end{align}
where we dropped terms of $\mathcal O (S^0)$, furthermore $\sigma^{\pm} (t) = e^{\pm i \Delta t}\sigma^{\pm}$ and $\sigma^z(t) = \sigma^z$. Using Eqs.~\eqref{appeq:chitk}~and~\eqref{appeq:chiok}, we can identify the susceptibility in each terms of the coupling in the form of $\chi_{nn}^\perp(\omega,k_x) = \tfrac i \hbar \!\int_0^\infty dt\, e^{-i(\omega-\varepsilon_{k_x,n}) t-\eta t}$. As it is shown in App.~\ref{app:SWdiss}, the linewidth $\eta$ can be replaced by the Gilbert damping $\hbar \Gamma(\varepsilon_{k_x,n}) = \aG \varepsilon_{k_x,n}$. Finally, we get
\begin{align}
\begin{split}
W_\text{eff} =& \frac 1 8 \sum_{p,q} \sum_{k_x,n} e^{ik_x x_{pq}} \vec{M}^-_{-k_x,n} \cdot \boldsymbol{\sigma}_q\\
&\times \left \lbrace \tfrac 1 2  M^{++}_{k_x,n} \sigma_p^- \chi_{nn}^\perp(\Delta/\hbar, k_x) + \tfrac 1 2 M^{+-}_{k_x,n} \sigma_p^+ \chi_{nn}^\perp(-\Delta/\hbar, k_x)\right.  \\
&\ \ \ \   \left.+ M^{+z}_{k_x,n} \sigma_p^z \chi_{nn}^\perp(0, k_x) \right \rbrace  + h.c.,
\end{split}
\label{appeq:Weffij}
\end{align}
where $x_{pq} = x_p - x_q$. Dropping the off-resonant terms $\chi_{nn}^\perp(-\Delta/\hbar, k_x)$ and $\chi_{nn}^\perp(0, k_x)$ and expanding $\vec{M}^-_{-k_x,n} \cdot \boldsymbol{\sigma}_q$ leads to Eq.~\eqref{eq:W12}. For ferromagnetic interlayer coupling, i.e., $J^\perp > 0$ (that is $\Delta < 0$), the $\chi_{nn}^\perp(-\Delta/\hbar, k_x)$   term becomes the resonant contribution. In this latter case the leading contribution to the coupling would be $\propto |M^{+-}|^2$ that is of the same order as the magnon-induced relaxation rate in the ferromagnetic coupling case [see Eq.~\eqref{appeq:relaxomega}]. This is a reason why in our work we focus on the  antiferromagnetic interlayer coupling.

We note that the off-resonant terms cannot be dropped for the qubit splitting corrections [i.e., the $p = q$ terms in Eq.~\eqref{appeq:Weffij}]. Considering $\Delta \approx \varepsilon_0$, where $\varepsilon_0$ is the energy of the chiral mode at $k_x = 0$,  the resonant term gives a contribution on the order of the coupling strength ($\sim 1\,$neV), and the $\chi_{nn}^\perp(0, k_x)$ term is expected to be even smaller. The contribution of $\chi_{nn}^\perp(-\Delta/\hbar, k_x)$ on the other hand contains terms of the order of $|M^{+-}|^2/(2\Delta)$ (second order in exchange) that are orders of magnitude stronger than the formers, i.e., $|M^{+-}| \sim 100\,\mu$eV, leading to a dynamical contribution to the effective field of the order of $1\,\mu$eV. Since this contribution is still a small corrections to the static exchange field one can simply redefine $\vec B_\text{eff}$ accordingly.

\subsection{Linear spectrum}

The chiral edge mode has a linear dispersion around $k_x \sim 0$ and it is well separated in energy from the bulk modes. Therefore the main contribution to the susceptibility at the corresponding energy range is given by 
\begin{align}
\chi_{nn}^\perp(\Delta/\hbar, k_x) \approx -2S\hbar \frac{\delta_{ne}}{v_x (k_x + i \kappa) - \delta}
\end{align}
where $\delta = \Delta - \varepsilon_0$ and $\kappa^{-1} \approx  \frac{v_x}{\aG \Delta}$ is the mean free path of the chiral magnon. The other chiral branch with opposite group velocity is localized on the other edge of the sample and therefore is neglected in the effective qubit-qubit coupling. 

First we convert the sum over $k_x$ to an integral as 
\begin{align}
\begin{split}
W_{pq} = \frac 1 {16}  \frac{a_x}{2\pi}\!\int \limits_{-\pi/a_x}^{\pi/a_x} \! dk_x\, e^{ik_x x_{pq}} M^{++}_{k_x,n} \sigma_p^- \chi_{ee}^\perp(\Delta/\hbar, k_x) \\
\times \vec{M}^-_{-k_x,n} \cdot \boldsymbol{\sigma}_q + h.c.,
\end{split}
\label{appeq:Wijintegral}
\end{align}
assuming that the sample is large enough, i.e., $2\pi /L_x \rightarrow 0$. If the integral is extended to infinity, i.e., $a_x \rightarrow 0$, it can be performed using the residue theorem. However, this approximation is only valid in the low-energy limit, or in our specific case for $|\delta| \ll |v_x/a_x| \sim J S$, such that the pole of the integrand remains at finite $k_x$. The integral of interest can be evaluated using residue theorem as
\begin{align}
\begin{split}
-\frac{S a_x}{\pi} \int \limits_{-\infty}^\infty dk_x& \frac{e^{ik_x x_{ij}}}{v_x (k_x + i \kappa) - \delta} f(k_x) \\
 &=\Theta (-v_x x_{ij}) \frac{2iS a_x}{|v_x|} e^{ik_0 x_{ij} - |\kappa x_{ij}|} f(k_0-i\kappa)\, ,
\end{split}
\label{appeq:Wijresidue}
\end{align} 
where $k_0 = \delta/v_x$, Furthermore, we have assumed that $f(k_x)$ is a holomorphic function and the contribution of the upper arc goes to zero if the contour is extended to infinity. Thus, for the two-qubit couplings ($p\neq q$) we have   
\begin{align}
&W_{12}+W_{21} = \frac{iS a_x}{8 |v_x|} e^{- |\kappa | d} M^{++}_{k_0,n}\vec{M}^-_{-k_0,n} \nonumber \\ 
&\hspace{20pt}\times \left[\Theta (v_x)e^{-ik_0 d} \sigma_1^-  \boldsymbol{\sigma}_2  + \Theta (-v_x) e^{ik_0 d} \sigma_2^- \boldsymbol{\sigma}_1\right]  + h.c.,
\end{align}
which can be rewritten as  
\begin{align}
&W_{12}+W_{21} = \frac{iS a_x}{16 |v_x|} e^{- |\kappa|d} (M^{++}_{k_0,n} M^{-+}_{-k_0,n} e^{-i|k_0|d} \sigma_1^-  \sigma^-_2 \nonumber  \\
&\hspace{20pt}+\text{sgn}(v_x)|M^{++}_{k_0,n}|^2 e^{-ik_0 d}\sigma_1^-  \sigma^+_2 +M^{++}_{k_0,n}M^{-z}_{-k_0,n}  \nonumber  \\
&\hspace{20pt} \times e^{-i|k_0|d}(\Theta (v_x) \sigma_1^- \sigma_2^z  + \Theta (-v_x) \sigma_1^z \sigma_2^-)  + h.c.
\end{align}
The individual two-qubit matrix elements can be read off directly to obtain Eq.~\eqref{eq:allW12coupschiral} of the main text.

\subsection{Quadratic spectrum}

Similarly to the case of the chiral edge magnon, we can discuss the effect of a topologically trivial magnonic mode that is localized at the edge of the FM. To this, we assume that the energy of the trivial mode $\varepsilon_{k_x,e} = \varepsilon_0 + D_x k_x^2$ is well separated from the two bulk bands and therefore the single mode approximation is adequate. The effective interaction matrix elements then read as
\begin{subequations}
\label{appeq:triveffcoup}
\begin{align}
\begin{split}
\bra{01} W_{12}\ket{10} 
\approx & -\frac{S a_x}{D_x}  \text{Re}\!\left[ \frac{e^{-K|d|}}{K} \right] |M^{++}_{-K,e}|^2 \, ,
\end{split}
\end{align}
\vspace{-.5cm}
\begin{align}
\begin{split}
\bra{11}  W_{12} \ket{00} 
  \approx -\frac{S a_x}{D_x K }  e^{-K|d|} M^{++}_{-K,e} (M^{+-}_{-K,e})^*\, ,
\end{split}
\end{align}
\vspace{-.5cm}
\begin{align}
\begin{split}
\bra{10}  W_{12} \ket{00} 
\approx -\frac{S a_x}{D_x K} e^{-K|d|}  M^{++}_{-K,e} (M^{+z}_{-K,e})^* \, ,
\end{split}
\end{align}
\vspace{-.5cm}
\begin{align}
\begin{split}
\bra{01} W_{12} \ket{00} = \bra{10} W_{12} \ket{00} \, ,
\end{split}
\end{align}
\end{subequations}
where $K = D_x^{-1/2} \sqrt{-\delta + i \aG \Delta}$ is a complex wave number, the real and imaginary parts of which describe the decay and the oscillations of the effective couplings, respectively. In Fig.~\ref{fig:triv_scales}, the real and imaginary parts of $K$ are shown as a function of the detuning, $\delta$. We see that below resonance ($\delta < 0$) the real part of $K$ is large and positive, leading to fast decay of the effective couplings in Eq.~\eqref{appeq:triveffcoup} as a function of qubit-qubit distance $d$, whereas the imaginary part becomes larger above resonance ($\delta > 0$) leading mostly to oscillations in the coupling strength.

Furthermore we note that the formulas above are only valid for $|K|\ll \pi/a_x$, or equivalently $\delta \ll D_x/a_x^2 \sim J S$. Therefore in the case of the bulk modes (harmonic spectrum) the exponential decay in the coupling is only valid close enough to the corresponding resonance. Consequently, the finite coupling in the middle of the gap is not captured by these analytic formulas. 

\begin{figure}
\includegraphics[width= 0.8\columnwidth]{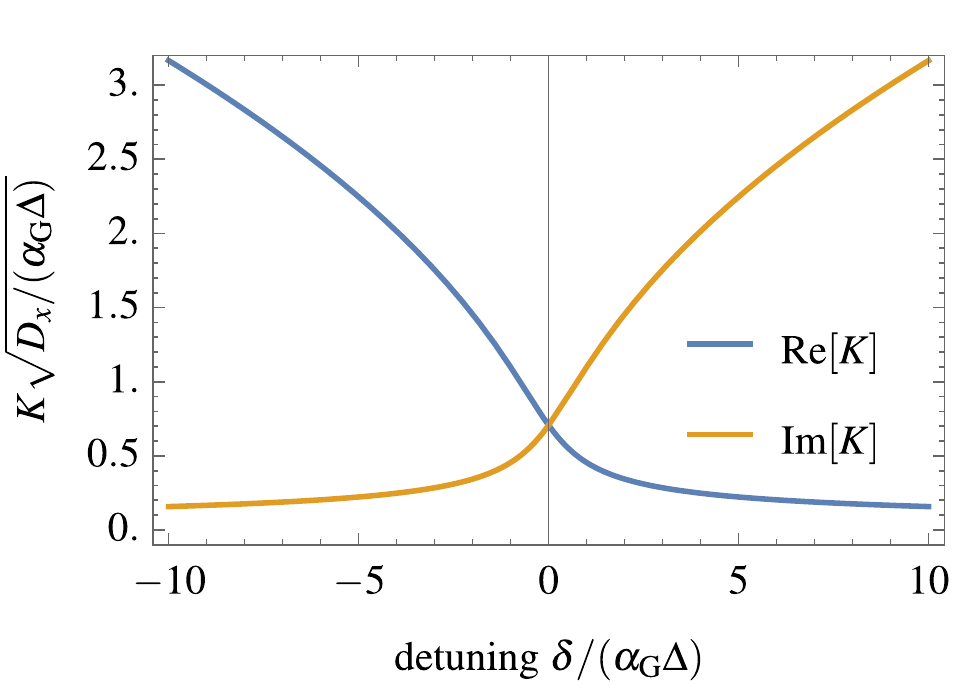}
\caption{Real and imaginary parts of the complex wave number $K = D_x^{-1/2} \sqrt{-\delta + i \aG \Delta}$, describing the decay and the period of the oscillations in the effective coupling of Eq.~\eqref{appeq:triveffcoup}, respectively.}
\label{fig:triv_scales}
\end{figure}

\section{Decoherence}
\label{app:decoh}
In this section we show how to relate the transversal $\NoSp_{V^-}$ and longitudinal $\NoSp_{V^z}$ qubit noise spectra to the transversal noise spectrum of the magnons.
 The transversal noise spectrum of magnons is defined as 
\begin{align}
\NoSp^\perp_{k_x,n} (\omega) \equiv \int \limits_{-\infty}^\infty dt\, e^{-i\omega t} \langle \{ S^-_{-k_x,n} (t) , S^+_{k_x,n} \} \rangle\, ,
\label{appeq:Sperpo}
\end{align}
substituting the time dependence of the FM spin operators in Eq.~\eqref{appeq:timeevoSkxn} the integral can be evaluated as 
\begin{align}
\NoSp^\perp_{k_x,n} (\omega) = 2\pi \hbar \delta(\hbar \omega - \varepsilon_{k_x,n}) \langle \{ S^-_{-k_x,n} , S^+_{k_x,n} \} \rangle\, , 
\label{appeq:Sperpdelta}
\end{align}
where we can replace $ 2\pi \hbar \delta(\hbar \omega - \varepsilon_{k_x,n})$ by $\tfrac 1 {S} \text{Im} [ \chi^\perp_{nn} (\omega, k_x) ]$ in the dissipative case. Furthermore, using the time dependence  of the FM spin operators in Eq.~\eqref{appeq:timeevoSkxn} and substituting it into Eq.~\eqref{appeq:Sperpo}, one can easily show that
\begin{align}
\NoSp^\perp_{k_x,n} (-\omega) = \int \limits_{-\infty}^\infty dt\, e^{-i\omega t} \langle \{ S^+_{k_x,n} (t) , S^-_{-k_x,n} \} \rangle\, .
\label{appeq:Sperpmo}
\end{align}
Exploiting the commutation relations between the magnon creation and annihilation operators one obtains
\begin{align}
\NoSp^\perp_{k_x,n} (\omega) = \text{Im} [ \chi^\perp_{nn}(\omega,k_x) ] \coth (\beta \varepsilon_{k_x,n}/2)  \, ,
\end{align}
as stated in the main text. We note that this is just a form of the well-known fluctuation-dissipation theorem.

Writing down the transversal qubit noise spectrum according to its definition as $\NoSp_{V^-} (\omega)= \int\! dt \Big \{ [ V^- (t) ]^\dagger, V^- (0)\Big \}\, \mathrm{e}^{-\mathrm{i}\omega t}$, where $V^-$ is the term multiplying $\sigma_p^+$ in Eq.~\eqref{appeq:qubittoFMcoupling}, leads to 
\begin{align}
\begin{split}
\NoSp_{V^-}(\omega) = \frac{1}{4} \sum_{k_x,n} \int \limits_{-\infty}^\infty\! dt\, e^{-i\omega t} &\left[ |M^{++}_{k_x,n}|^2 \langle \{ S^-_{-k_x,n} , S^+_{k_x,n} \} \rangle\right. \\
&\left. + |M^{-+}_{k_x,n}|^2 \langle \{ S^+_{k_x,n} (t) , S^-_{-k_x,n} \} \rangle \right],
\end{split}
\end{align}
where we exploited that $\langle \{ S^-_{-k_x,n}, S^+_{k_x',n'} \} \rangle \propto \delta_{k_x,k_x'} \delta_{nn'}$. In the equation above, the perpendicular magnon noise spectum appears in the form of Eqs.~\eqref{appeq:Sperpo}~and~\eqref{appeq:Sperpmo}. Finally we get
\begin{align}
&\NoSp_{V^-}(\omega) = \frac{1}{4} \sum_{k_x,n}  \coth (\beta \varepsilon_{k_x,n}/2) \left[ |M^{++}_{k_x,n}|^2 \text{Im} [ \chi^\perp_{nn}(\omega,k_x) ] \right. \nonumber\\
&\hspace{30pt}\left. + |M^{-+}_{k_x,n}|^2 \text{Im} [ \chi^\perp_{nn}(-\omega,k_x) ] \right]\, ,
\label{appeq:relaxomega}
\end{align}
where the second term can be dropped for $\omega = \Delta/\hbar$ as it is strongly suppressed even in the dissipative case. An analogous derivation leads to the longitudinal qubit noise spectrum as
\begin{align}
\begin{split}
\NoSp_{V^z}(\omega) = \sum_{k_x,n}  & \coth (\beta \varepsilon_{k_x,n}/2)|M^{+z}_{k_x,n}|^2 \\
&\times\left[  \text{Im} [ \chi^\perp_{nn}(\omega,k_x) ]+ \text{Im} [ \chi^\perp_{nn}(-\omega,k_x) ] \right]\, .
\end{split}
\label{appeq:SVz}
\end{align}

Afterwards, the noise power spectra obtained in Eqs.~\eqref{appeq:relaxomega} and \eqref{appeq:SVz} can be used to obtain the decoherence rates in Bloch-Redfield approximation as 
\begin{subequations}
\begin{equation}
\Gamma_1 = \tfrac 1 { 4 \hbar^2}\NoSp_{V^-} (\Delta/\hbar)\, ,
\label{appeq:G1rel}
\end{equation}
\begin{equation}
\Gamma_2^* = \tfrac 1 {4 \hbar^2}\NoSp_{V^z} (0)\, ,
\label{appeq:G2deph}
\end{equation}
\end{subequations}
where $\Gamma_1$ is the qubit relaxation rate and $\Gamma_2^*$ is called pure dephasing.

\subsection{Decoherence due to a quadratic magnon mode}

\begin{figure}[!b]
\includegraphics[width= 0.8\columnwidth]{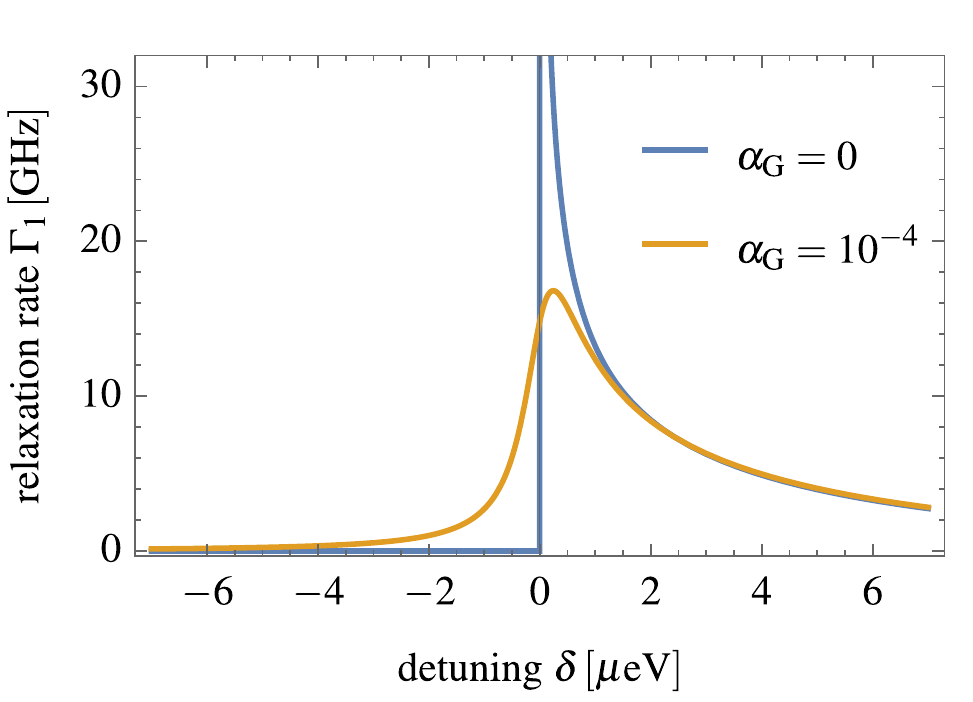}
\caption{Relaxation rate $\Gamma_1$  from Eq.~\eqref{appeq:G1} plotted as a function detuning $\delta$. The relaxation is caused by to a trivial (1D) edge magnon for $D_x = 0.5\,$meV$\cdot$nm$^2$. The effect of the van Hove singularity at zero detuning is smoothed out by the finite Gilbert damping.}
\label{fig:triv_rel}
\end{figure}

In the main text we focused on the decoherence rates due to the resonant interaction with the chiral magnon mode. Here we provide analogous formulas for the case of a quadratic mode, e.g., a bulk mode or a topologically trivial edge mode. We start the discussion with the non-dissipative limit $\aG =0$, where the noise spectrum of the edge mode assumes the form $\NoSp^\perp_{k_x,e} (\omega) \propto \delta (D_x k_x^2\!+\! \varepsilon_{0}\! -\! \hbar\omega)$ [see Eq.~\eqref{appeq:Sperpdelta}], and the decoherence rates of Eqs.~\eqref{appeq:G1rel}-\eqref{appeq:G2deph} become 
\begin{subequations}
\begin{align}
\Gamma_1 =& \Theta (\delta) \frac{S a_x}{2\hbar D_x k_0} \left( |M^{++}_{k_0,e}|^2+  |M^{++}_{-k_0,e}|^2\right)\, ,
\label{appeq:G1}
\end{align}
\begin{align}
\Gamma^*_2 \sim \mathcal O (S^0)\, ,
\label{appeq:G2}
\end{align}
\end{subequations}
where we used Eqs.~\eqref{appeq:relaxomega} and \eqref{appeq:SVz} and substituted them into Eqs.~\eqref{appeq:G1rel}-\eqref{appeq:G2deph}. Furthermore, $k_0 = \sqrt{\delta/D_x}$ and we assumed that the FM spectrum is gapped (e.g., via external magnetic field) and therefore $\NoSp^\perp_{k_x,e} (0) = 0$. The divergent behaviour at $k_0 = 0$ is due to the van Hove singularity of the density of states that can be observed in Fig.~\ref{fig:triv_rel}.

In order to account for the effect of Gilbert damping we assume $d_y,\, d_z\ll l_x$ and neglect the dependence of the coupling on $k_x$ except for the Gaussian factor $e^{-k_x^2l_x^2/2}$. In contrast to Eqs.~\eqref{appeq:G1} and \eqref{appeq:G2}, we consider the imaginary part of the susceptibility with quadratic dispersion in Eq.~\eqref{appeq:chiok} with a finite linewidth $\eta = \aG \varepsilon_{k_x,e}$, the integral over momentum is exactly solvable leading to 
\begin{subequations}
\begin{align}
\Gamma_1 =& -\frac{S a_x}{\hbar D_x} |M^{++}_{0,e}|^2\, \text{Im}\! \left[ \frac{e^{K^2l_x^2/2}\,  \text{Erfc}(\tfrac{K l_x}{\sqrt 2})}{K} \right] ,
\end{align}
\begin{align}
\Gamma^*_2 \approx \frac{\aG}{2\!\sqrt{\pi}} \frac{a_x}{l_x} \frac{|M^{-z}_{0,F}|^2}{\hbar\Delta_F} \left( 1 + \frac{D_F}{l_x^2 \Delta_F}  \right)\, ,
\end{align}
\end{subequations}
where we listed the relaxation rate for a trivial edge mode, assuming $|M^{++}_{0,e}|^2 \sim 100\, \mu$eV for the coupling matrix element, and the dephasing rate for the FM resonance mode. Furthermore, $\Delta_F$ is the ferromagnetic resonance energy and $D_F$ is the curvature of the lowest magnonic band at $k_x=0$. Even though the damping smoothens out the divergence of the density of states at resonance, the relaxation rate is still highly enhanced (see Fig.~\ref{fig:triv_rel}) rendering the trivial edge mode unfavourable in practical applications.

\section{Exchange interaction --- Analytical formulas}
\label{app:ex}

In this section we provide details of the direct exchange-induced FM-QD coupling and derive the effective analytical formula presented in Eq.~\eqref{eq:MkxJ} of the main text. Since the QD layer is adjacent to the FM layer, the wavefunction of the particle on the QD can have a finite overlap with the FM spins. In real space the interaction can be written as
\begin{align}
-\sum_{i, \mu} \vec S_{i,\mu} \cdot \mathbf{\hat J}_{i,\mu} \boldsymbol \sigma |\psi (\vec r_i + \vec r^\mu -  \vec r_\text{\tiny{QD}})|^2\,.
\end{align}
Assuming that the interlayer exchange interaction $\mathbf{\hat J}_i$ is homogeneous and isotropic with a strength of $J^\perp$ and keeping the leading terms only in the $1/S$ expansion we get
\begin{align}
-S J^\perp \sigma^z + \frac{1}{4} \sum_{k_x,n} (S_{-k_x,n}^+ M^{-+}_{k_x,n} \sigma^- + S_{-k_x,n}^- M^{+-}_{k_x,n}\sigma^+) \,.
\end{align}
Here, the first term provides the effective magnetic field as $\vec B_\text{eff} = -\tfrac 1 {\mu_B}SJ^\perp \vec e_z$ as well as we get 
\begin{align}
\begin{split}
M^{-+}_{k_x,n} = -2J^\perp \sum_{i,\mu}& e^{-ik_x x_i} \varphi^\mu_{-k_x,n} (y_i)\\ 
& \times |\psi (x_i + x^\mu ,y_i + y^\mu -  y_\text{\tiny{QD}})|^2.
\end{split}
\label{appeq:Mex}
\end{align}
In what follows, the QD wave function is assumed to be Gaussian in both spatial directions, i.e.,
\begin{align}
|\psi (x_i + x^\mu ,y_i + y^\mu)|^2 = \frac{a_x a_y}{4\pi l_x l_y} e^{-(x_i+x^\mu)^2l_x^{-2}} e^{-(y_i+y^\mu)^2l_y^{-2} } \, . \label{appeq:discrWF}
\end{align}

Next we derive the estimate for the coupling to the chiral edge mode in the continuum approximation. To this end we convert the sum over $x$ to an integral and evaluate it as
\begin{align}
\frac{1}{\sqrt \pi l_x} \int \limits_{-\infty}^\infty dx\, e^{-i k_x x} e^{-(x+x^\mu)^2l_x^{-2}}  = e^{-k_x^2 l_x^2/4 + ik_x x^\mu}\, .\label{appeq:FTQDWF}
\end{align}
The coupling matrix element then reads as
\begin{align}
M^{-+}_{k_x,e} \approx -\frac{J^\perp a_y}{2\sqrt \pi l_y}e^{-k_x^2 l_x^2/4} \sum_{y_i,\mu} e^{-ik_x x^\mu} \varphi^\mu_{-k_x,e} (y_i) e^{-(y_i+y^\mu - d_y)^2l_y^{-2} }.
\end{align}
Next, we exploit that  the edge state is well localized around $y_i \sim 0$ and neglect $x^\mu \ll l_x$ and $y^\mu \ll l_y$ in the formulas to arrive at 
\begin{align}
M^{-+}_{k_x,e} \approx -\frac{J^\perp a_y}{2\sqrt \pi l_y}e^{-k_x^2 l_x^2/4} e^{- d_y^2l_y^{-2} } \sum_{y_i,\mu} \varphi^\mu_{-k_x,e} (y_i) \, .
\end{align}
Finally, since $k_x \lesssim l_x^{-1} \ll \pi/a_x$ the last sum can be replaced by $\delta \mu_e$ leading to Eq.~\eqref{eq:MkxJ} of the main text.

We can also account for the exponential envelope of the edge mode and arrive at
\begin{align}
M^{-+}_{k_x,e} \approx -\frac{J^\perp a_y}{4\lambda}e^{-k_x^2 l_x^2/4} e^{- d_y/\lambda + l_y^2/(4\lambda^2)} \left[1+\erf \left( \tfrac{l_y}{2\lambda} - \tfrac{d_y}{l_y}\right)\right]\, .
\label{appeq:MkxJerf}
\end{align}
This approximation is necessary to capture the qualitative dependence of the coupling for large $d_y$ due to the short-ranged nature of the direct exchange interaction.

\section{Dipole-dipole interaction analytical formulas}
\label{app:dip}

In this section we provide details of the dipole-induced FM-QD coupling and derive the effective analytical formula presented in Eqs.~\eqref{eq:dipdip} and \eqref{eq:Mkx0ky} of the main text. The dipole-dipole interaction between localized magnetic moments reads as
\begin{align}
\begin{split}
H_{d-d} = &-\frac{\mu_0}{4\pi} \frac{3(\vec m_1 \cdot \vec{\hat r})(\vec m_2 \cdot \vec{\hat r})-\vec m_1 \cdot \vec m_2}{|\vec r_1 - \vec r_2|^{3}} 
\\
&+ \mu_0 \frac 2 3 \vec m_1 \cdot \vec m_2\, \delta (\vec r_1 - \vec r_2 ) \, ,
\end{split}
\label{appeq:dipcl}
\end{align}
where the magnetic moments are $\vec m_1 = -\mu_B g \vec S_i$ with $\vec S_i$ being the FM spin at position $\vec r_1$ and $\vec m_2 =  -\mu_B  g_\text{\tiny{QD}} |\psi(\vec r_2)|^2 \boldsymbol \sigma$, and we define $\vec{\hat r} = (\vec r_1 - \vec r_2)/|\vec r_1 - \vec r_2|$. Using the wavefunction of the QD, $\psi(x',y')$, given in Eq.~\eqref{appeq:discrWF}, the coupling between the QD and a lattice spin at position $(x,y)$ is given by 
\begin{subequations}
\label{appeq:Mdip}
\begin{align}
\begin{split}
M^{-+}(x,y) = -&\frac{\mu_0 \mu_B^2 g g_\text{\tiny{QD}}}{4\pi} \sum_{x',y'} |\psi (x',y')|^2\\
& \times \frac{(x-x')^2+(y-y')^2-2d_z^2}{[(x-x')^2+(y-y')^2+d_z^2]^{5/2}}\, ,
\end{split}
\end{align}
\begin{align}
\begin{split}
M^{--}(x,y) = -&\frac{\mu_0 \mu_B^2 g g_\text{\tiny{QD}}}{4\pi} \sum_{x',y'} |\psi (x',y')|^2\\ 
& \times \frac{3[(x-x')-i(y-y')]^2}{[(x-x')^2+(y-y')^2+d_z^2]^{5/2}}\, ,
\end{split}
\end{align}
\begin{align}
\begin{split}
M^{-z}(x,y) = -&\frac{\mu_0 \mu_B^2 g g_\text{\tiny{QD}}}{4\pi} \sum_{x',y'} |\psi (x',y')|^2\\ 
& \times \frac{3[(x-x')-i(y-y')]d_z}{[(x-x')^2+(y-y')^2+d_z^2]^{5/2}}\, ,
\end{split}
\end{align}
\end{subequations}
where $d_z$ is the distance between the QD and the FM planes. Furthermore we note that $M^{+-}(x,y) = M^{-+}(x,y)$,  $M^{++}(x,y) = [M^{--}(x,y)]^*$, and  $M^{+z}(x,y) = [M^{-z}(x,y)]^*$. 

Using the couplings in Eq.~\eqref{appeq:Mdip}, we write the coupling of the QD to a given magnon mode $(k_x,n)$ as
\begin{align}
\begin{split}
\vec M^-_{k_x,n} = \sum_{i,\mu} e^{-ik_x x_i} \varphi^{\mu}_{-k_x,n}(y_i) \sum_{i', \mu'} |\psi(\vec r'_i + \vec r^{\mu'})|^2\\
\times \vec D^-(\Delta x, \Delta y, d_z)\, ,
\end{split}
\label{appeq:Mkxndip}
\end{align}
where we have used Eq.~\eqref{appeq:MxytoMkxn}. Furthermore we defined
\begin{align}
\begin{split}
\vec D^- (\Delta x, \Delta y, d_z)=&-\frac{\mu_0 \mu_B^2 g g_\text{\tiny{QD}}}{4\pi (\Delta x^2 + \Delta y^2+ d_z^2)^{5/2}} \\
&\times 
\begin{pmatrix}
\Delta x^2 + \Delta y^2 - 2d_z^2\\
3(\Delta x - i \Delta y)^2\\
3(\Delta x + i\Delta y)d_z
\end{pmatrix}
\begin{matrix}
\, +\\
\,  -\\
\, z
\end{matrix}\ ,
\end{split}
\end{align}
with  $\Delta x = x_i - x'_i + x^\mu - x^{\mu'}$ and $\Delta y = y_i - y'_i + y^\mu - y^{\mu'}$. Note that $\vec D^-$ is not given in a vector form, but the first (second, third) element of the column correspond to the $D^{-+}$ ($D^{--}$, $D^{-z}$) coupling elements. This notation is emphasized next to the corresponding row.

In order to obtain the formulas for the edge mode in the continuum approximation we convert the sums over $x$ coordinates to integrals, switch to the center-of-mass frame, and neglect $x^\mu, x^{\mu'}\sim a_x$ since the QD wavefunction changes slowly on this scale. Furthermore, we make use of Eq.~\eqref{appeq:FTQDWF} to get
\begin{align}
\begin{split}
&\sum_{x_i,x_i'} e^{-i k_x x_i} |\psi(x'_i)|^2 \vec D^-(\Delta x, \Delta y, d_z) \\
&=e^{-k_x^2l_x^2/4} \frac 1 {a_x}\int \limits_{-\infty}^{\infty} dx\, e^{-ik_x x} \vec D^-(x, \Delta y, d_z)\, ,
\end{split}
\end{align}
where the Fourier transformation of the $\vec D^-$ is analytically solvable and reads as 
\begin{align}
\begin{split}
&\frac 1 {a_x}\int \limits_{-\infty}^{\infty} dx\, e^{-ik_x x} \vec D^-(x, \Delta y, d_z)=-\frac{\mu_0 \mu_B^2 g g_\text{\tiny{QD}}}{4\pi a_x}\\
& \times \begin{pmatrix}
-\tfrac 2 3 k_x^2 K_0 + \tfrac 2 3 \left|\tfrac{k_x}{d_\perp}\right| K_1 + \tfrac 2 3 \tfrac{k_x^2}{d_\perp^2} (\Delta y^2 - 2d_z^2)K_2\\
- 2 k_x^2 K_0 + 2 \left|\tfrac{k_x}{d_\perp}\right| (1- 2k_x \Delta y) K_1 + \tfrac 2 3 \tfrac{k_x^2\Delta y^2}{d_\perp^2} K_2\\
-2id_z \left(  \left|\tfrac{k_x}{d_\perp}\right| k_x K_1 + \tfrac{k_x^2}{d_\perp^2} \Delta y K_2 \right)
\end{pmatrix}
\begin{matrix}
\, +\\
\,  -\\
\, z
\end{matrix}\ ,
\end{split}
\label{appeq:Dkxgen}
\end{align}
with $K_n \equiv K_n(|k_x| d_\perp)$ being the $n$th modified Bessel functions of the second kind, and $d_\perp = \sqrt{\Delta y^2 + d_z^2}$.

\begin{figure}[t]
    \includegraphics[width= \columnwidth]{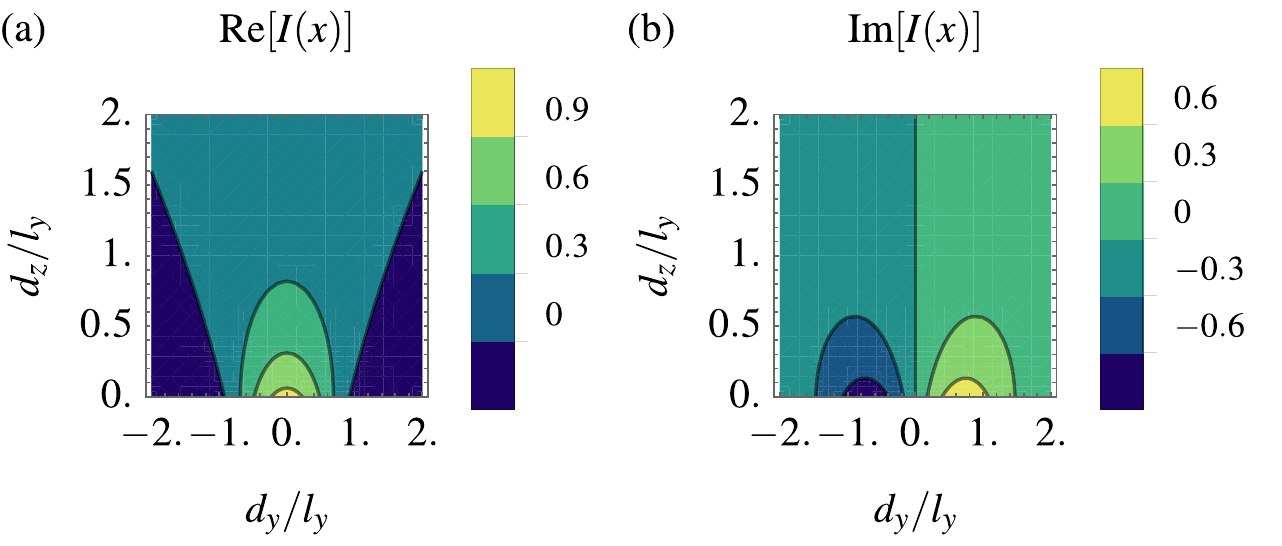}
    \caption{ (a) Real and (b) imaginary part of the function $I(x)$ in Eq.~\eqref{appeq:Iofx} with $x = (id_y - d_z)/l_y $. These functions determine the dependence of the dipolar FM-QD coupling of Eq.~\eqref{appeq:M0edip} on the relative length scales $d_y/l_y$ and $d_z/l_y$. } 
    \label{appfig:Mdip}
\end{figure}

From this point onwards, we will be focusing on the $k_x \sim 0$ case since $l_x \gg d_y,d_z, l_y$. In this case, Eq.~\eqref{appeq:Dkxgen} for $k_x = 0$   simplifies to
\begin{align}
\begin{split}
\frac 1 {a_x}\int \limits_{-\infty}^{\infty} dx\, \vec D^-(x, \Delta y, d_z)=\frac{\mu_0 \mu_B^2 g g_\text{\tiny{QD}}}{2\pi a_x d_\perp^4} \begin{pmatrix}
d_z^2-\Delta y^2\\
\Delta y^2-d_z^2\\
2i\Delta y d_z
\end{pmatrix}
\begin{matrix}
\, +\\
\,  -\\
\, z
\end{matrix}\ .
\end{split}
\label{appeq:Dkx0}
\end{align}
Assuming $y_i,y'_i \ll l_y$ we substitute Eq.~\eqref{appeq:Dkx0} back into Eq.~\eqref{appeq:Mkxndip} to get 
\begin{align}
\begin{split}
\vec M^-_{k_x\sim 0,e} =& \frac{\mu_0 \mu_B^2 g g_\text{\tiny{QD}}}{2\pi a_x} e^{-k_x^2l_x^2/4} \sum_{y_i,\mu} \varphi^\mu_{0,e}(y_i) \\
&\times \frac{a_y}{\sqrt \pi l_y}\sum_{y'_i} e^{-(y'_i-d_y^2)l_y^{-2}} \tfrac{1}{d_\perp^4} \begin{pmatrix}
d_z^2-\Delta y^2\\
\Delta y^2-d_z^2\\
2i\Delta y d_z
\end{pmatrix}
\begin{matrix}
\, +\\
\,  -\\
\, z
\end{matrix}\ .
\end{split}
\end{align}
Once again we assume that the edge mode is well localized around $y_i \sim 0$ and therefore $\Delta y = y'_i$. Then the sum over $y'_i$ can be converted to an integral and coupling acquires its final form
\begin{align}
\begin{split}
\vec M^-_{k_x\sim 0,e} =&\frac{\mu_0 \mu_B^2 g g_\text{\tiny{QD}}}{\pi a_x l_y^2} e^{-k_x^2l_x^2/4}\delta \mu_e\\
& \times \begin{pmatrix}
1+ \sqrt \pi \text{Re}[x e^{x^2} (1+\erf (x)) ]\\
-1- \sqrt \pi \text{Re}[x e^{x^2} (1+\erf (x)) ]\\
-i \sqrt \pi \text{Im}[x e^{x^2} (1+\erf (x)) ]
\end{pmatrix}
\begin{matrix}
\, +\\
\,  -\\
\, z
\end{matrix}\ ,
\end{split}
\label{appeq:M0edip}
\end{align}
where $x = (id_y - d_z)/l_y $ and the formula is valid for $k_x^{-1} \gg l_y, d_y, d_z$, if the QD covers several lattice sites i.e., $l_x, l_y \gg a$ and the edge mode is very well localized e.g., $l_y \gg \lambda$. Furthermore we note that for $d_z \lesssim a $ the coupling starts to depend on the lattice structure $\vec r^\mu$ that we have neglected in the calculation above. Finally, we introduce the complex function
\begin{align}
I(x) = 1 + \sqrt \pi  x \mathrm{e}^{x^2} [1+\text{erf}(x)]\, ,
\label{appeq:Iofx}
\end{align}
in order to simplify the formula for the dipole interaction-induced couplings $M^{-+}_{0,e}$, $M^{--}_{0,e}$, and $M^{-z}_{0,e}$ in Eq.~\eqref{eq:dipdip} of the main text. The real and imaginary parts of $I(x)$ determine the dependence of the coupling matrix elements $\vec M^-_{0,e}$ on the relative length scale $x = (id_y - d_z)/l_y$. This functional dependence is shown in Fig.~\ref{appfig:Mdip}.

Lastly, we show the derivation of the dipole coupling to the FM resonance mode deep in the bulk, where the magnonic eigenmodes can be labelled by the quantum numbers $k_x$ and $k_y$. The final result for this coupling has been shown in Eq.~\eqref{eq:Mkx0ky} of the main text. Starting from
\begin{align}
\begin{split}
M^{-z}_{k_x,k_y} = \sum_{x_i,y_i} e^{-ik_x x_i -ik_y y_i} \sum_{i', \mu'}& |\psi(\vec r'_i + \vec r^{\mu'})|^2\\
& \times D^{-z}(\Delta x, \Delta y, d_z)\, ,
\end{split}
\end{align}
we separate center-of-mass and relative coordinates obtaining the Gaussian factors $e^{-k_x^2l_x^2/4} e^{-k_y^2l_y^2/4}$ from the center-of-mass integrals. Considering $k_x = 0$ in the relative coordinates we are left with the integral
\begin{align}
    \frac{1}{a_x a_y}\int  dx dy\, e^{-ik_y y} D^{-z}(x, y, d_z)= \frac{\mu_0 \mu_B^2 g g_\text{\tiny{QD}}}{2 a_x a_y}  e^{-|k_y|d_z} k_y\, ,
\end{align}
which together with the Gaussian factors yield Eq.~\eqref{eq:Mkx0ky}. Finally we note that the $g$-tensor anisotropy in the QD layer can be taken into account as in Eq.~\eqref{appeq:dipolefield} for the dipole field, by replacing $g_\text{\tiny{QD}} \boldsymbol \sigma$ with $\vec{ \hat g} \boldsymbol \sigma$, where $\vec{ \hat g}$ is the $g$-tensor of the QD. 

\bibliography{references}

\end{document}